\documentclass[12pt,a4paper]{article}
\pdfoutput=1
\usepackage{graphicx}
\usepackage{amsfonts}
\usepackage{pdftexcmds}
\usepackage{color}
\usepackage{amssymb}
\usepackage{amsmath}

\usepackage{amsthm}
\newtheorem*{lemma*}{Lemma}

\numberwithin{equation}{section}
\date{}

\newcommand{\diff}{\mathrm{d}}

\title{\bf {Casimir pistons with generalized boundary conditions: a step forward}}
\author{Guglielmo Fucci$^1$\footnote{fuccig@ecu.edu}, Klaus Kirsten$^{2,3}$\footnote{klaus\_kirsten@baylor.edu} and Jose M. Mu$\tilde{\rm n}$oz-Casta$\tilde{\rm n}$eda$^4$\footnote{jose.munoz.castaneda@uva.es}\\
\footnotesize{{\sl $^1$Department of Mathematics, East Carolina University, Greenville, NC 27858 USA.}}\\
\footnotesize{{\sl $^2$GCAP-CASPER Department of Mathematics, Baylor University, Waco, TX 76798, USA.}}\\
\footnotesize{{\sl $^3$Mathematical Reviews, American Mathematical Society, 416 4th St., Ann Arbor, MI 48103, USA}}\\
\footnotesize{{\sl $^4$Departamento de F\'isica te\'orica, at\'omica y \'optica, Universidad de Valladolid, Valladolid, SPAIN}}}

\begin{document}

\maketitle
\begin{center}
\date{\today}
\end{center}

\begin{abstract}
In this work we study the Casimir effect for massless scalar fields propagating in a piston geometry of the type $I\times N$ where $I$ is an interval of the
real line and $N$ is a smooth compact Riemannian manifold. Our analysis represents a generalization of previous results obtained for pistons configurations
as we consider all possible boundary conditions that are allowed to be imposed on the scalar fields. We employ the spectral zeta function formalism
in the framework of scattering theory in order to obtain an expression for the Casimir energy and the corresponding Casimir force on the piston.
We provide explicit results for the Casimir force when the manifold $N$ is a $d$-dimensional sphere and a disk.

\end{abstract}


\section{Introduction}

The Casimir effect is undoubtedly one of the most interesting physical phenomena predicted by quantum field theory. Since the seminal work of
Casimir in 1948 \cite{casimir}, interest on the subject, and more generally on the influence that external conditions have on a quantum system, has steadily increased.
In fact the literature regarding the Casimir effect has grown not only in the number of works produced but also in its scope. When it was first theoretically predicted in
\cite{casimir}, the Casimir effect focused simply on the attraction between two perfectly conducting neutral plates. Since then the Casimir effect has
been studied for a plethora of different geometric configurations, quantum systems and boundary conditions (see for instance \cite{bordag09,bordag01,milton01,plunien86}
and references therein for a review on the subject). One of the most interesting and widely analyzed geometric configurations is
the piston geometry which was first introduced by Calvalcanti in \cite{caval04}. While one can find a number of specific piston configurations throughout the
literature (\cite{barton06,edery06,edery07,edery08,hertz05,hertz07,kirsten09,li97,marachevsky07,milton09,morales10} represents a, necessarily incomplete, list of examples),
the most general one can be described as consisting of two compact manifolds, referred to as chambers, possessing a common boundary of co-dimension one representing the piston.

The reason for the widespread interest enjoyed by piston configurations lies mainly in the following important feature: In general calculations of the Casimir energy
for quantum systems propagating in a given geometric configuration and subject to suitable boundary conditions lead, by the very nature of the phenomenon,
to divergent quantities. In this case one is confronted with the non-trivial task of extracting, from these divergent results, meaningful physical information about the
Casimir effect. In the case of piston configurations these problems are somewhat mitigated. In fact, while the Casimir energy of pistons might be divergent, the Casimir
force acting on the piston itself is, is many instances, a well-defined quantity. In this regard, it is worth pointing out that piston configurations with non-vanishing
curvature can have a divergent Casimir force acting on the piston \cite{fucci11b,fucci11,fucci12}. The Casimir force acting on a piston depends not only on the specific
geometry of the piston configuration but also on the boundary conditions that are imposed on the quantum field. In fact the Casimir force acting on a piston of a
specific geometry can vary substantially as the boundary conditions are changed. For this reason, a precise and comprehensive analysis of the influence that
the boundary conditions have on the Casimir force is of paramount importance for a deeper understanding of the Casimir effect. Studying the effect that boundary conditions have
on the Casimir force on pistons is not only of theoretical significance but it could also shed some light on the Casimir effect of quantum systems consisting of real, as
opposed to idealized, materials. In fact, suitable boundary conditions can be utilized to describe physical properties of real materials. Some results regarding
the Casimir effect with general boundary conditions have been obtained, for instance, in \cite{aso-mc-npb13} for the case of parallel plates and in
\cite{fucci-npb15,fucci16,fucci-ijmpa17} in regards to piston configurations. It is important to mention, for completeness, that real materials could be
modeled by smooth potentials with compact support rather than boundaries (see e.g. \cite{acto95,beau13,fucci11c,fucci16a,eliz97}).

This work is mainly aimed at generalizing the results, obtained in \cite{fucci-npb15,fucci-ijmpa17}, for the Casimir effect in piston configurations. We consider a piston
configuration of the type $I\times N$ where $I\subset\mathbb{R}$ is a closed interval of the real line and $N$ is a smooth compact Riemannian manifold with or without
boundary $\partial N$. We analyze a massless scalar field propagating in the aforementioned geometric configuration endowed with the most general boundary conditions for which  the Laplace operator describing its dynamics admits strongly consistent self-adjoint extensions. It is important to emphasize, at this point, that the results presented in this
paper for the Casimir energy and corresponding force on the piston encompass \emph{all possible} boundary conditions that can be imposed on scalar fields propagating on pistons of the type $I\times N$, and, hence,
represent an exhaustive analysis of the Casmir effect for scalar fields propagating on these types of pistons. In order to perform
such general analysis we exploit the results obtained in \cite{asorey05} which enable one to characterize all self-adjoint extensions of the Laplacian.
By following the techniques employed in \cite{mukibo-lmp15}, we will utilize spectral zeta function regularization methods in order to derive explicit expressions
for the desired Casimir energy and the corresponding force on the piston. We perform the analysis of the spectral zeta function of the piston configuration by
relying primarily on methods from scattering theory. While there are other methods to obtain the spectral zeta function of the system under consideration,
we are of the opinion that the formalism based on scattering theory provides a somewhat more transparent physical interpretation of our results.

The outline of the paper is as follows. In the next section we describe in detail the piston configuration and the general boundary conditions to
be imposed on the scalar field. Subsequently, we utilize scattering methods in order to obtain an integral representation of the spectral zeta function.
We then analytically continue the representation and derive an expression for the Casimir energy and corresponding force on the piston for the piston
under consideration. In the last sections we present some particular cases as examples of our general results. The conclusions provide a summary of our main results and
some ideas for possible further studies in this area.

\section{Generalities about the quantum vacuum.}\label{sec2new}
For the sake of completeness we include a brief description of the main physical ideas underlying the scalar Casimir energies and forces. Typically the quantum field Hamiltonian that governs the dynamics of a quantum scalar field without self-interaction is given by
\begin{equation}
	\mathbb{H}_{_{QFT}}=\sum_{\omega}\hbar\omega\left(\widehat N_\omega+\frac{1}{2} \right),
\end{equation}
where the set $\{\omega\}$ are the normal modes of the quantum scalar field, and $\widehat N_\omega$ is the number operator for the particular frequency $\omega$. As can be seen, a quantum scalar field without self-interaction is nothing but a grand canonical ensemble of non relativistic quantum harmonic oscillators. The frequencies $\{\omega\}$ of the collection of harmonic oscillators are determined by the spectrum of a Schr\"odinger operator $\widehat K$. Specifically, the frequencies are the positive square roots of the eigenvalues of the operator $\widehat K$, i. e. $\{\omega\}=\{\omega \in\mathbb{R}^+\,\,\text{such that}\,\,\omega^2\in\sigma(\widehat K)\}$, being $\sigma(\widehat K)$ the spectrum of the operator $\widehat  K$. In order to have a well-defined quantum field theory, such that the quantum Hamiltonian $	\mathbb{H}_{_{QFT}}$ is self-adjoint, all the frequencies $\{\omega\}$ of the quantum field's normal modes must be real and non negative. Hence the Schr\"odinger operator $\widehat K$ must be self-adjoint and non-negative, to ensure that $	\mathbb{H}_{_{QFT}}$ is self-adjoint.

The vacuum of a quantum field theory defined under the conditions mentioned above, is the state in which there are no particles for any of the modes of the field, i. e. all the harmonic oscillators are in their fundamental state. Hence, if we denote the vacuum by $\vert \boldsymbol{0}\rangle$, the expectation for the number operators in this state is
$$
\langle \boldsymbol{0}\vert \widehat N_\omega\vert\boldsymbol{0}\rangle=0 ,
$$
which immediately enables one to obtain a formal expression for the quantum vacuum energy, defined as the expectation value of the quantum field Hamiltonian for the vacuum state:
\begin{equation}
	E_0=\langle \boldsymbol{0}\vert 	\mathbb{H}_{_{QFT}}\vert\boldsymbol{0}\rangle=\sum_{\omega^2\in\sigma(\widehat K)}\frac{\hbar\omega}{2}.\label{qvacE}
\end{equation}
As it usually happens in quantum field theory this expression is divergent, because the Schr\"odinger operator $\widehat K$ is non-negative and unbounded. This does not mean that the expression in Eq. \eqref{qvacE} is meaningless. To extract a physically meaningful quantity out of this last expression we need to regularize and renormalize the quantum vacuum energy. There are many regularization methods that are useful for different situations, that have been used in the last 30 years (see Ref. \cite{bordag09} for a review). In this paper we will use the zeta function regularization. We can regularize the quantum vacuum energy in Eq. \eqref{qvacE} by introducing a dimensionless complex parameter $s$ and a regularization mass $\mu$, to rewrite the regularized quantum vacuum energy as
\begin{eqnarray}
	E_0(s)= &=&\frac{\mu^{2 (s+1/2)}}{2}\hbar\sum_{\omega^2\in\sigma(\widehat K)}\omega^{-2 s}
= \frac{\mu^{2 (s+1/2)}} 2 \hbar\,\zeta_{\widehat K} (s),\label{qvacE2}
\end{eqnarray}
where $\mu$ is a regularization parameter with dimension of a mass, and $\zeta _{\widehat K} (s)$ is the zeta function 
associated with the operator $\widehat K$, namely
$$\zeta_{\widehat K} (s) = \sum_{\omega^2 \in \sigma (\widehat K)} \omega^{-2s}.$$
The formal quantum vacuum energy in Eq. (\ref{qvacE}) corresponds to $s=-1/2$. Typically, $\zeta_{\widehat K} (s)$ has the following expansion about 
$s=-1/2$,
\begin{eqnarray}
\zeta_{\widehat K} \left(-\frac 1 2 +\epsilon \right) = \frac 1 \epsilon \mbox{Res } \zeta_{\widehat K} \left( - \frac 1 2 \right) + \mbox{FP } \zeta_{\widehat K} \left( - \frac 1 2 \right) + {\cal O } (\epsilon ), \label{lauzeta}
\end{eqnarray}
where $\mbox{Res}$ denotes the residue and FP stands for the finite part at $\epsilon =0$, which corresponds to the ${\cal O} (\epsilon ^0)$-term in this Laurent expansion.
One then has
\begin{eqnarray}
E_0 \left( -\frac 1 2 + \epsilon \right)= \frac 1 {2\epsilon} \mbox{Res } \zeta_{\widehat K} \left( - \frac 1 2 \right) 
+ \frac 1 2 \left[ \mbox{FP } \zeta_{\widehat K} \left( - \frac 1 2 \right) + \mbox{Res } \zeta _{\widehat K} \left(- \frac 1 2 \right) \ln \mu^2 \right]
+ {\cal O} (\epsilon ) , \label{vaclau}
\end{eqnarray}
and defines
\begin{eqnarray}
E_{Cas} = \frac 1 2 \left[ \mbox{FP } \zeta_{\widehat K} \left( - \frac 1 2 \right) + \mbox{Res } \zeta _{\widehat K} \left(- \frac 1 2 \right) \ln \mu^2 \right], \label{vacren}
\end{eqnarray}
indicating that the Casimir energy generally has finite ambiguities that are proportional to $\mbox{Res } \zeta _{\widehat K} (-1/2 )$.
For massive fields an extra renormalization condition can be imposed to remove the regularization dependence. On the other hand, for massless fields this regularization dependence can not be removed in general.

\paragraph{The quantum vacuum force.}Despite the possible regularization dependence of the quantum vacuum energy for the case of massless scalar fields, the physical quantity that is measured in a laboratory to detect the quantum vacuum energy is the force it does produce between two macroscopic objects. It has been demonstrated that the part of the quantum vacuum energy that contains the distance between two macroscopic objects is independent of the regularization parameter even for massless fields (see e. g. Refs. \cite{emig07,KK}). This result holds in particular for piston systems like the case we will study in this paper because $\mbox{Res } \zeta_{\widehat K} ( -  1/ 2 )$ for the relevant $\widehat K$ will not depend on the position of the piston.

To finish this section, we would like to remark that henceforth we set $\hbar =1$.

\section{The general setup: $U(4)$ boundary conditions}
We begin our analysis by considering a direct product manifold $M$ of the type $M=I\times N$. In this setting we define $I=[0,L]\subset\mathbb{R}$ to be a closed interval of the real line and $N$ to be a smooth compact $d$-dimensional Riemannian manifold with or without a boundary $\partial N$. It is clear from the above definition that $M$ has
dimension $D=d+1$. The piston configuration can be obtained from the manifold $M$ following the construction detailed in \cite{fucci-npb15,fucci-ijmpa17}. The two
chambers of the piston are realized by dividing the manifold $M$ with a cross-sectional manifold $N_{a}$ at the point $a\in(0,L)$.
The manifold $N_{a}$ represents the piston itself. The two chambers $M_{I}$ and $M_{II}$ are, by construction, smooth compact $D$-dimensional Riemannian
manifolds with boundary $\partial M_{I}=N_{0}\cup N_{a}\cup ([0,a]\times\partial N)$ and $\partial M_{II}=N_{a}\cup N_{L}\cup ((a,L]\times\partial N)$, respectively.

Let $\psi(t,x)$ with $x\in M$ denote a massless scalar field propagating on the piston configuration outlined above, and $\phi(x)$ denote the normal modes in which the scalar field decomposes after writing down the Fourier mode decomposition for $\psi(t,x)$ in the time coordinate.
Due to the direct product structure of $M$ we can write the equation characterizing the normal modes of the scalar field $\phi$ as the eigenvalue equation
\begin{equation}\label{0}
-\left(\frac{\diff^{2}}{\diff x^2}+\Delta_{N}\right)\phi=\alpha^2\phi\;,
\end{equation}
where $\Delta_{N}$ denotes the Laplacian on the manifold $N$. By using separation of variables we can write the solution $\phi$ as the product of a longitudinal
part and a cross-sectional one, namely $\phi=f(x)Y(\Omega)$ where $x$ is the coordinate in the interval $I$ and $\Omega$ denotes the coordinates on $N$. The functions
$Y(\Omega)$ are eigenfunctions of the operator $\Delta_{N}$ with eigenvalue $\lambda$
\begin{equation}\label{1}
-\Delta_{N}Y(\Omega)=\lambda^2 Y(\Omega)\;,
\end{equation}
while $f(x)$ satisfies the simple second-order differential equation in the space ${\cal I}=[0,a)\cup (a,L]$
\begin{equation}\label{2}
-\frac{\diff^2}{\diff x^2}f_{\lambda}(x,k)=k^{2}f_{\lambda}(x,k)\;,
\end{equation}
where, for notational convenience, we have introduced the parameter $k^{2}=\alpha^2-\lambda^2$. The parameter $k^2$ becomes the eigenvalue once the differential
equation (\ref{2}) is augmented by appropriate boundary conditions. As previously stated, we will consider all possible boundary conditions that can be
imposed on $f_{\lambda}(x,k)$
which lead to a self-adjoint boundary value problem. According to the methods developed in \cite{asorey05} this is equivalent to considering all possible
non-negative self-adjoint extensions of the operator in (\ref{2}). We would like to point out that we will consider only strongly consistent
self-adjoint extensions of the operator in (\ref{2}), that is all the self-adjoint extensions that are non-negative independently on the size of the interval
$I$ \cite{mukibo-lmp15}. We would like to mention at this point that the theory of self-adjoint extensions of Sturm-Liouville operators on two  intervals is well-known and can be found, for instance, in \cite{zettl} and references therein. However, in this work we choose to utilize a formalism to describe all the relevant self-adjoint extensions that is rooted in the physical language \cite{asorey05}. In addition, we will be using quantum mechanical scattering theory which does offer a clearer and much more immediate physical interpretation of the formal discussions that will follow.

\subsection{General boundary conditions: $U(4)$}
The boundary of ${\cal I}$ consists of four points, namely $\partial{\cal I}=\{x=0, x=a^-,x=a^+,x=L \}$. According to the formalism developed in \cite{asorey05,aso-mc-npb13,mukibo-lmp15}, the boundary conditions that characterize a given self-adjoint extension of the differential operator in (\ref{2}) are expressed in the following form
\begin{equation}\label{3}
\varphi-i\dot{\varphi}=U(\varphi+i\dot{\varphi})\;,
\end{equation}
where  $\varphi$ is a vector with entries being the boundary values of the function $f_{\lambda}(x,k)$, and $\dot{\varphi}$ denotes a vector whose entries are the outgoing
normal derivative of $f_{\lambda}(x,k)$ at the boundary (c.f. \cite{aso-mc-npb13,mukibo-lmp15}), that is
\begin{equation}\label{4}
\varphi=\left(
\begin{array}{c}
 f_{\lambda}(0,k) \\
 f_{\lambda}\left(a^-,k\right) \\
 f_{\lambda}\left(a^+,k\right) \\
 f_{\lambda}(L,k) \\
\end{array}
\right);\,\,
\dot{\varphi}=\left(
\begin{array}{c}
 -f_{\lambda}^\prime(0,k) \\
 f_{\lambda}^\prime\left(a^-,k\right) \\
 -f_{\lambda}^\prime\left(a^+,k\right) \\
 f_{\lambda}^\prime(L,k) \\
\end{array}
\right)\Rightarrow \varphi\pm i\dot{\varphi}=\left(\begin{array}{c}
 f_{\lambda}(0,k)\mp i f_{\lambda}^\prime(0,k) \\
 f_{\lambda}\left(a^-,k\right)\pm if_{\lambda}^\prime\left(a^-,k\right) \\
 f_{\lambda}\left(a^+,k\right) \mp i f_{\lambda}^\prime\left(a^+,k\right)\\
 f_{\lambda}(L,k)\pm if_{\lambda}^\prime(L,k) \\
\end{array}
\right)\equiv\Psi^\pm\;.
\end{equation}
Since the set of self-adjoint extensions of the differential operator in (\ref{2}) defined over ${\cal I}$ is in
one-to-one correspondence with the elements of the group $U(4)$ \cite{asorey05}, the matrix $U$ in (\ref{4}) must be an element
of the unitary group $U(4)$. This means that for any given $U\in U(4)$ we obtain a corresponding self-adjoint extension of the second derivative operator
in (\ref{2}) defined on the domain \cite{mukibo-lmp15}
\begin{equation}\label{5}
{\cal D}_{U}=\{f_{k}(x)\in H^{2}([0,L],\mathbb{C}): \varphi-i\dot{\varphi}=U(\varphi+i\dot{\varphi})\}\;,
\end{equation}
which is a subspace of the Sobolev space $H^{2}([0,L],\mathbb{C})$. It must be noted, that not all self-adjoint extensions give rise to a well-defined quantum field theory. Taking into account the fact that the normal modes of the scalar massless quantum field confined in the piston are characterized by the non-relativistic Schr\"odinger eigenvalue problem \eqref{0}, only those self-adjoint extensions that are non-negative can be used to construct a meaningful scalar quantum field theory on the piston.

In order to explicitly implement the boundary conditions (\ref{3}) we need a solution of the differential equation (\ref{2}) which
can be easily found to be of the form
\begin{equation}\label{6}
f_\lambda(x,k)=\begin{cases}
 A_1 e^{i k x}+B_1 e^{-i k x} & 0\leq x\leq a^- \\
 A_2 e^{i k x}+B_2 e^{-i k x} & a^+\leq x\leq L\;,
\end{cases}
\end{equation}
where the constants $\{A_1, B_1, A_2, B_2\}$ are to be determined as to satisfy the boundary conditions and the normalization condition.
By using the explicit solution (\ref{6}), the boundary vectors $\Psi^{(\pm)}$ defined in (\ref{4}) are given by
\begin{equation}\label{7}
\Psi^{(\pm)}=\left(
\begin{array}{c}
 (1\pm k) A_1+(1\mp k) B_1 \\
 e^{i a k} (1\mp k) A_1+e^{-i a k} (1\pm k) B_1 \\
 e^{i a k} (1\pm k) A_2+e^{-i a k} (1\mp k) B_2 \\
 e^{i k L} (1\mp k ) A_2+e^{-i kL} (1\pm k) B_2 \\
\end{array}\right)=M_{\pm}\cdot \left(
\begin{array}{c}
 A_1 \\
 B_1 \\
 A_2 \\
 B_2 \\
\end{array}
\right)\;,
\end{equation}
where we have introduced the following matrix
\begin{equation}\label{8}
M_{\pm}\equiv \left(
\begin{array}{cccc}
 1\pm k & 1\mp k & 0 & 0 \\
 e^{i a k} (1\mp k) & e^{-i a k} (1\pm k) & 0 & 0 \\
 0 & 0 & e^{i a k} (1\pm k) & e^{-i a k} (1\mp k) \\
 0 & 0 & e^{i k L} (1\mp k) & e^{-i k L} (1\pm k) \\
\end{array}
\right)\;.
\end{equation}
By substituting (\ref{7}) into \eqref{3} we obtain the homogeneous linear system
\begin{equation}\label{9}
(M_{-}-U\cdot M_{+})\cdot \left(
\begin{array}{c}
 A_1 \\
 B_1 \\
 A_2 \\
 B_2 \\
\end{array}
\right)=0\;.
\end{equation}
The above linear system has a non-trivial solution for the parameters $\{A_1, B_1, A_2, B_2\}$ if and only if the determinant of the matrix $(M_{-}-U\cdot M_{+})$
vanishes, that is
\begin{equation}\label{10}
\det (M_{-}-U\cdot M_{+})=0\;.
\end{equation}
This expression represents an equation for the parameter $k$ whose solutions determine the eigenvalues of the boundary value problem consisting of the
differential equation \eqref{2} and the boundary conditions associated with $U\in U(4)$. In order to obtain an explicit expression for (\ref{10})
we need an appropriate representation of a generic element $U\in U(4)$. One way of proceeding is to notice that $U(4)\cong (SU(4)\times U(1))/\mathbb{Z}_{4}$ and , hence,
an element $U\in U(4)$ can be written as $U=e^{i\theta}\bar{U}$ where $\theta\in[0,2\pi]$ and $\bar{U}\in SU(4)$ which, in turn, can be represented
in terms of Euler angles and $4\times 4$ Gell-Mann-type matrices as shown in \cite{tilma02}.
Since $\textrm{dim}(U(4))=16$, the relation (\ref{10}) would contain sixteen free real parameters. Although with the help of a computer algebra program one
could in principle obtain an explicit expression for (\ref{10}) in terms of the required free parameters, it is, in our opinion, more instructive to consider
simpler cases. Indeed, the large number of free parameters to follow by considering the full $U(4)$ would certainly obfuscate the main physical properties of
the quantum system which represent the focus of our work.

To this end, starting with the next section, we will restrict our attention to boundary conditions that are represented by matrices belonging to
the direct product subgroup $U(2)\times U(2)\subset U(4)$.

\section{$U(2)\times U(2)$ reductions and topology change}

The restriction to the subset $U(2)\times U(2)$ of $U(4)$ allows us to analyze the most general boundary conditions that relate pairs of boundary points of ${\cal I}$.
If we denote ${\bf U}_1\in U(2)$ and ${\bf U}_2\in U(2)$ as
\begin{equation}
{\bf U}_1=\begin{pmatrix}
 a_{11}& a_{12} \\
  a_{21}& a_{22}\\
\end{pmatrix}\;, \quad \textrm{and} \quad
{\bf U}_2=\begin{pmatrix}
 b_{11}& b_{12} \\
  b_{21}& b_{22}\\
\end{pmatrix}\;,
\end{equation}
then a generic element of $U(2)\times U(2)\subset U(4)$ describing boundary conditions that relate pairs of boundary points of ${\cal I}$ have one of the following forms:
\begin{equation}\nonumber
V=\left(
\begin{array}{cccc}
  a_{11}& a_{12} & 0& 0 \\
  a_{21}& a_{22}& 0& 0 \\
  0 & 0& b_{11}& b_{12} \\
 0 & 0& b_{21}& b_{22}\\
\end{array}
\right)\;, \quad W=\left(
\begin{array}{cccc}
 a_{11} & 0 & 0  & a_{12} \\
 0 & b_{11} & b_{12} & 0 \\
 0 & b_{21} & b_{22} & 0 \\
 a_{21} & 0 & 0  & a_{22} \\
\end{array}
\right)\;,
\end{equation}
\begin{equation}\label{11}
R=\left(
\begin{array}{cccc}
  a_{11}& 0 & a_{12}& 0 \\
  0 & b_{11}& 0& b_{12} \\
  a_{21} & 0& a_{22}& 0 \\
 0 & b_{21} & 0 & b_{22}\\
\end{array}
\right)\;.
\end{equation}
It is not very difficult to realize that each of the matrices displayed in (\ref{11}) characterizes a specific class of boundary conditions.

Boundary conditions described by matrices of the form $V$ in (\ref{11}) couple the boundary conditions at $x=0$ and $x=a^-$ through a $U(2)$ matrix
and the boundary conditions at $x=a^+$ and $x=L$ through, in general, another $U(2)$ matrix. In more detail, by using $V$ in (\ref{11}) in the relation (\ref{3}) we get
\begin{equation}\label{12}
\left(\begin{array}{c}
f_{\lambda}(0,k)+if_{\lambda}^\prime(0,k) \\
f_{\lambda}(a^-,k)-if_{\lambda}^\prime(a^-,k)
\end{array}\right)
={\bf U}_1\left(\begin{array}{c}
f_{\lambda}(0,k)-if_{\lambda}^\prime(0,k) \\
f_{\lambda}(a^-,k)+if_{\lambda}^\prime(a^-,k)
\end{array}\right)\;,
\end{equation}
\begin{equation}\label{12a}
\left(\begin{array}{c}
f_{\lambda}(a^+,k)+if_{\lambda}^\prime(a^+,k) \\
f_{\lambda}(L,k)-if_{\lambda}^\prime(L,k)
\end{array}\right)
={\bf U}_2\left(\begin{array}{c}
f_{\lambda}(a^+,k)-if_{\lambda}^\prime(a^+,k) \\
f_{\lambda}(L,k)+if_{\lambda}^\prime(L,k)
\end{array}\right)\;.
\end{equation}
This case represents two disconnected chambers, since
the quantum vacuum fluctuations in one chamber are independent from the ones in the other chamber. The spectrum of the boundary value problem
(\ref{2}) and (\ref{3}) is given, in this situation, simply by the union of the spectra of the self-adjoint extension defining the dynamics in each of the chambers.
The disconnected chamber configuration has been already covered in \cite{fucci-npb15} and, hence, will not be discussed further in this work.

Matrices of the form $W$ in (\ref{11}) describe, instead, the case in which the boundary conditions at $x=0$ and $x=L$ are coupled through a $U(2)$ matrix
and the boundary conditions at $x=a^-$ and $x=a^+$ are coupled, generally, through another $U(2)$ matrix. That is, the condition (\ref{3}) becomes,
\begin{equation}\label{13}
\left(\begin{array}{c}
f_{\lambda}(0,k)+if_{\lambda}^\prime(0,k) \\
f_{\lambda}(L,k)-if_{\lambda}^\prime(L,k)
\end{array}\right)
={\bf U}_1\left(\begin{array}{c}
f_{\lambda}(0,k)-if_{\lambda}^\prime(0,k) \\
f_{\lambda}(L,k)+if_{\lambda}^\prime(L,k)
\end{array}\right)\;,
\end{equation}
\begin{equation}\label{13a}
\left(\begin{array}{c}
f_{\lambda}(a^-,k)-if_{\lambda}^\prime(a^-,k) \\
f_{\lambda}(a^+,k)+if_{\lambda}^\prime(a^+,k)
\end{array}\right)
={\bf U}_2\left(\begin{array}{c}
f_{\lambda}(a^-,k)+if_{\lambda}^\prime(a^-,k) \\
f_{\lambda}(a^+,k)-if_{\lambda}^\prime(a^+,k)
\end{array}\right)\;,
\end{equation}
which can easily be obtained by replacing $U$ in (\ref{3}) with $W$ defined in (\ref{11}). In this case, as it is clear from (\ref{13a}), the quantum fluctuations are allowed to travel through the piston itself, a situation which occurs when the piston is not opaque. Using the boundary conditions (\ref{13a}) is equivalent to modeling the piston itself as a point supported potential.
This case would complement the analysis of semi-transparent pistons \cite{morales10} and pistons with transmittal boundary conditions \cite{fucci-ijmpa17}. The boundary conditions in (\ref{13}) can induce a topology change as they allow for the two ends $x=0$ and
$x=L$ of the piston configuration to be identified. In this case the piston configuration would have the topology of a torus.

Matrices of the form $R$ in (\ref{11}) characterize the situation in which boundary conditions at $x=0$ are coupled, through a $U(2)$ matrix, to the boundary
conditions at $x=a^{+}$ while boundary conditions at $x=a^{-}$ are coupled to the ones at $x=L$ through another $U(2)$ matrix. Although formally this case
leads to a boundary value problem which is strongly self-adjoint, it is not suitable for describing a piston configuration. In fact, fields propagating in the left chamber
would be constrained by the boundary at $x=0$ but would have no constraints on the right boundary of that chamber, namely $x=a^-$.
This leads to a scenario which would \emph{de facto} eliminate the left chamber since fields propagating in it would ``\emph{feel}" the left boundary but not the right one.
A similar argument applies to the fields propagating in the right chamber since, in this case, the piston itself is completely opaque.
Because of the remarks above, we will be focusing our analysis on the membrane configuration.

\subsection{Membrane configuration}

There are basically two approaches that can be applied to the analysis of the membrane configuration.
The first consists of writing a solution of the differential equation (\ref{2}) as a linear combination of sine and cosine functions and
then impose the boundary conditions in (\ref{13})-(\ref{13a}). The second approach, instead, is based on the formalism of scattering theory where the solutions are written
in terms of transmission and reflection amplitudes. In the analysis that will follow we use the latter approach since, in our opinion, it describes
the Casimir effect for the membrane configuration in a physically more meaningful way.
To carry out the calculation we will follow a procedure consisting of two steps:
\begin{enumerate}
\item We start by studying the piston wall over the entire real line. In this case the piston wall can be described as a potential supported at the point $x=a$ defined by the boundary conditions \eqref{13a} through the unitary matrix ${\bf U}_2$. The scattering states obtained in this case will satisfy \eqref{13a} independently of the presence of the external walls at $x=0,L$.

\item Afterwards we built the quantum field normal modes as linear combinations of the previously found scattering states and impose, on them, the boundary condition \eqref{13}, given by the unitary matrix ${\bf U}_1$, at the external points of the piston $x=0,L$.

\end{enumerate}
With this approach we can characterize the spectrum of normal modes of the massless quantum scalar field in terms of non-relativistic scattering data of the piston wall. This characterization enables one to have a better intuition about the phenomena appearing in the Casimir force in terms of the physical properties of the piston that are encoded in the scattering data.
To this end, we express the eigenfunctions of \eqref{2} with the boundary conditions (\ref{13}) and (\ref{13a}) as the following linear combination
\begin{equation}
f_\lambda(x,k)=A_{\lambda}(k)\psi_{\lambda,k}^{R}(x; {\bf U}_2)+B_{\lambda}(k) \psi_{\lambda,k}^{L}(x; {\bf U}_2)\;,\label{ans-pist-wf}
\end{equation}
where $\psi_{\lambda,k}^{R}(x; {\bf U}_2)$ and  $\psi_{\lambda,k}^{L}(x; {\bf U}_2)$ are the left-to-right and the right-to-left scattering states, respectively, and should be determined by the boundary condition  (\ref{13a}). On the other hand the coefficients $A_{\lambda}(k)$ and $B_{\lambda}(k)$ are determined by the boundary condition (\ref{13}).

\subsubsection{The piston on the real line}

The functions $\psi_{\lambda,k}^{R}(x; {\bf U}_2)$ and  $\psi_{\lambda,k}^{L}(x; {\bf U}_2)$ are solutions to the scattering problem
consisting of a point supported potential, positioned at $x=a$,
described by the unitary matrix ${\bf U}_2$. According to standard scattering theory, the left-to-right ($\psi_{\lambda,k}^R(x; {\bf U}_2 )$) and the right-to-left ($\psi_{\lambda,k}^L(x; {\bf U}_2 )$) scattering states can be written as
\begin{equation}
\psi_{\lambda,k}^R(x; {\bf U}_2)=\begin{cases}
 e^{-i k x} \tilde{r}_R+e^{i k x} & -\infty<x<a \\
 e^{i k x} \tilde{t}_R & a<x<\infty
\end{cases};\quad \psi_{\lambda,k}^L(x; {\bf U}_2 )=\begin{cases}
 e^{-i k x} \tilde{t}_L & -\infty<x<a \\
 e^{i k x} \tilde{r}_L+e^{-i k x} & a<x<\infty\;.
\end{cases}\label{ans-sc-disp}
\end{equation}
In order to determine the scattering data $\{\tilde{t}_R,\tilde{r}_R,\tilde{t}_L,\tilde{r}_L\}$ we impose the boundary conditions (\ref{13a}) on the
functions $\psi_{\lambda,k}^{R}(x; {\bf U}_2)$ and $\psi_{\lambda,k}^{L}(x; {\bf U}_2)$ separately.

By using $\psi_{\lambda,k}^R(x; {\bf U}_2 )$ in (\ref{13a}) we obtain
\begin{equation}\label{14}
\left(\begin{array}{c}
e^{-2ika}\tilde{r}_{R}(1-k)+(1+k) \\
\tilde{t}_{R}(1-k)
\end{array}\right)
={\bf U}_2\left(\begin{array}{c}
e^{-2ika}\tilde{r}_{R}(1+k)+(1-k) \\
\tilde{t}_{R}(1+k)
\end{array}\right)\;.
\end{equation}
An explicit expression for the linear system that determines the coefficients $\tilde{r}_{R}$ and $\tilde{t}_{R}$ can be found
by exploiting the Euler parametrization for ${\bf U}_2$,
that is
\begin{equation}\label{15}
{\bf U}_2=e^{i \theta } \left[\mathbb{I} \cos (\gamma )+i \sin (\gamma ) \left(q_1 \sigma _1+q_2
   \sigma _2+q_3 \sigma _3\right)\right]\;,
\end{equation}
where $\sigma_j$ represents the Pauli matrices, $(q_{1},q_{2},q_{3})$ is a unit vector $q_{1}^{2}+q_{2}^{2}+q_{3}^{2}=1$, and
$\theta\in [-\pi,\pi]$ and  $\gamma\in [-\pi/2,\pi/2]$. The solution to the linear system (\ref{14}) with the parametrization (\ref{15})
can then be written as $\tilde{t}_{R}=t_{R}$ and $\tilde{r}_{R}=e^{2ika}r_{R}$ where $t_{R}$ and $r_{R}$ are the scattering amplitudes for the case in which the piston is located at $x=0$
\begin{equation}\label{16}
t_R=\frac{-2 i k \left(q_1-i q_2\right) \sin (\gamma )}{D_{{\bf U}_{2}}(k)}\;,\quad r_R=\frac{\left(k^2+1\right) \cos (\gamma )+\left(k^2-1\right) \cos (\theta )+2 i k q_3
   \sin (\gamma )}{D_{{\bf U}_{2}}(k)}\;,
\end{equation}
where we have introduced the function
\begin{equation}
D_{{\bf U}_{2}}(k)=\left(k^2+1\right) \cos (\theta)+\left(k^2-1\right) \cos (\gamma )+2 i k \sin
   (\theta )\;\label{Usec}.
\end{equation}

\begin{center}
\begin{figure}[htbp]
\centerline{\includegraphics[width=10cm]{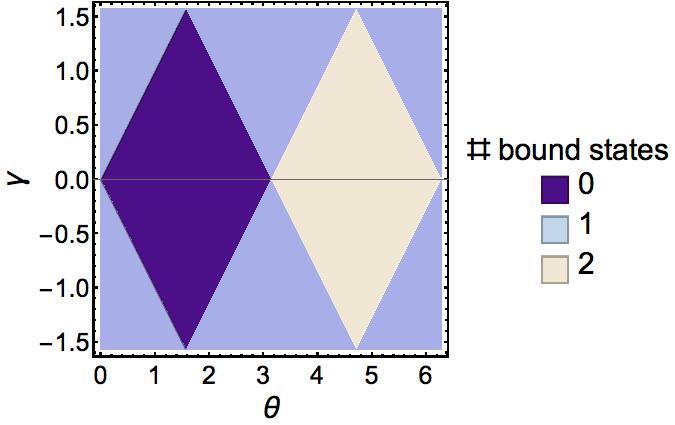}}
\caption{\small Bound states distribution in the $\theta-\gamma$ plane.
\label{fig1}}
\end{figure}
\end{center}

By imposing the boundary conditions (\ref{13a}) to the right-to-left scattering state $\psi_{\lambda,k}^{L}(x; {\bf U}_2)$
one finds a linear system for the coefficients $\tilde{t}_L$ and $\tilde{r}_L$ similar to the one in (\ref{14}). By using the parametrization
(\ref{15}) one can write the solutions for the right-to-left scattering coefficients as $\tilde{t}_{L}=t_{L}$ and $\tilde{r}_{L}=e^{-2ika}r_{L}$
where
\begin{equation}\label{17}
t_L=\frac{-2 i k \left(q_1+i q_2\right) \sin (\gamma )}{D_{{\bf U}_{2}}(k)}\;,\quad r_L=\frac{ \left(k^2+1\right) \cos (\gamma )+\left(k^2-1\right) \cos (\theta )-2 i k q_3
   \sin (\gamma )}{D_{{\bf U}_{2}}(k)}\;.
\end{equation}
We would like to point out that the scattering coefficients in $\psi_{\lambda,k}^{R}(x; {\bf U}_2)$ and $\psi_{\lambda,k}^{L}(x; {\bf U}_2)$ do satisfy the usual relations
$|r_{R}|^2+|t_{R}|^2=1$ and $|r_{L}|^2+|t_{L}|^2=1$, which imply, in particular, that the function $D_{{\bf U}_{2}}(k)$ cannot vanish for real $k>0$.
However, it is possible for $D_{{\bf U}_{2}}(k)$ to have zeroes on the positive imaginary $k$-axis. In fact, the solutions of the equation $D_{{\bf U}_{2}}(i\kappa)=0$
with $\kappa>0$ determine the bound states of the system \cite{galindo}. The solutions can be found to be
\begin{equation}
\kappa_{\pm}=-\tan\left(\frac{\theta\pm\gamma}{2}\right)\;.
\end{equation}
Since $\theta\in [-\pi,\pi]$ and  $\gamma\in [-\pi/2,\pi/2]$, it is not difficult to realize that it is possible to have either no bound states, one bound state, or
two bound states. The regions in the $\theta-\gamma$ plane leading to no, one, or two bound states is given in Figure \ref{fig1}.
It is important to notice, that in order to have a unitary quantum field theory all the normal modes of the field must have real non-negative frequencies. This means, in particular, that the scattering problem we have just analyzed can not have bound states. Hence we have to restrict ourselves to those unitary matrices ${\bf U}_2$ that give rise to non negative self-adjoint extensions, i. e. the dark purple zone in Fig. \ref{fig1}.

\subsubsection{The confined piston}
The eigenfunctions $f_{\lambda}(x,k)$ in (\ref{ans-pist-wf}) automatically satisfy the boundary conditions on the piston itself when
we use $\psi_{\lambda,k}^{R}(x; {\bf U}_2)$ and $\psi_{\lambda,k}^{L}(x; {\bf U}_2)$ in (\ref{ans-sc-disp}) with the coefficients found in (\ref{16}) and (\ref{17}).
Our next task therefore is to impose the remaining boundary conditions on $f_{\lambda}(x,k)$, namely the ones at the edges $x=0$ and $x=L$ of the piston configuration.
The scattering states $\psi_{\lambda,k}^{R}(x; {\bf U}_2)$ and $\psi_{\lambda,k}^{L}(x; {\bf U}_2)$ allow us to write the column vector of the boundary data
of $f_{\lambda}(x,k)$ in (\ref{13})
as
\begin{equation}\label{18}
\left(
\begin{array}{c}
 f_\lambda (0,k)\pm f_\lambda^\prime(0,k) \\
 f_\lambda (L,k)\mp i f_\lambda^\prime(L,k)  \\
\end{array}
\right)=\tilde M_{\pm}\left(
\begin{array}{c}
 A_{\lambda}(k) \\
 B_{\lambda}(k) \\
\end{array}
\right) ,
\end{equation}
where we have defined the matrices
\begin{equation}\label{19}
\tilde M_{\pm}=\left(
\begin{array}{cc}
1\mp k \left(1-\tilde r_R\right)+\tilde r_R & (1\pm k) \tilde t_L \\
 e^{i k L} (1\pm k) \tilde t_R & e^{-i k L} \left((1\mp k)+e^{2 i k L}
   (1\pm k) \tilde r_L\right) \\
\end{array}
\right) .
\end{equation}
With this notation the boundary condition \eqref{13} reads
\begin{equation}
(\tilde M_+-{\bf U}_1\tilde M_-)\left(
\begin{array}{c}
 A_{\lambda}(k) \\
 B_{\lambda}(k) \\
\end{array}
\right)=\left(
\begin{array}{c}
 0 \\
 0 \\
\end{array}
\right) .\label{bc-u1-ab}
\end{equation}
In order for \eqref{bc-u1-ab} to have non-trivial solutions, the determinant of the coefficients of the linear system must vanish, that is
one obtains the secular equation
\begin{equation}\label{19a}
F_{\lambda}(k,a;S,{\bf U}_1):=\det(\tilde M_+-{\bf U}_1\tilde M_-)=0\;.
\end{equation}

The last condition represents an equation in the variable $k$ whose solutions provide, through the relation $\alpha^2=k^2+\lambda^2$,
the eigenvalues of the problem (\ref{2}) with boundary conditions
(\ref{13}) and (\ref{13a}). By utilizing (\ref{19a}) and after some lengthy but straightforward calculations one obtains an explicit expression for
$F_{\lambda}(k,a;S,{\bf U}_1)$ as follows
\begin{eqnarray}\label{20}
&&F_{\lambda}(k,a;S,{\bf U}_1)=e^{-ikL}\left[ C_{{\bf U}_1}^++k^2C_{{\bf U}_1}^--2k(1-\det({\bf U}_1))\right]\nonumber\\
&&-e^{ikL}\det(S)\left[ C_{{\bf U}_1}^++k^2C_{{\bf U}_1}^-+2k(1-\det({\bf U}_1))\right]\nonumber\\
&&+(r_Re^{ik(2a-L)}+r_Le^{-ik(2a-L)})\left[C_{{\bf U}_1}^+-k^2C_{{\bf U}_1}^-\right]\nonumber\\
&&+2k(r_Re^{ik(2a-L)}-r_Le^{-ik(2a-L)})(u_{11}-u_{22})+4k(u_{21}t_R+u_{12}t_L) ,\label{1f}
\end{eqnarray}
where $u_{ij}$ are the entries of the matrix ${\bf U}_1$, and we have defined, for any $2\times 2$ matrix ${\bf m}$, the quantities
 \begin{equation}\label{21}
C_{\bf m}^\pm=C_{\bf m}(\pm 1)=1+\det({\bf m})\mp{\rm tr}({\bf m})\;,
\end{equation}
which are nothing but the characteristic polynomial $C_{\bf m}(x)$ of ${\bf m}$ evaluated at $x=\pm 1$.
It is clear from (\ref{20}) that the function $F_{\lambda}(k,a;S,{\bf U}_1)$ depends explicitly on the unitary matrix ${\bf U}_1\in U(2)$, and on the matrix ${\bf U}_2\in U(2)$ through the scattering matrix for the point supported potential described by the unitary matrix ${\bf U}_{2}$
\begin{equation}\label{22}
S(k;{\bf U}_2)=\left(
\begin{array}{cc}
 \tilde t_R & \tilde r_L \\
 \tilde r_R & \tilde t_L \\
\end{array}
\right).
\end{equation}
The determinant of the matrix in (\ref{22}) can be computed by using the expressions in (\ref{16}) and (\ref{17}) of the scattering coefficients.
One finds explicitly
\begin{equation}\label{23}
\det(S)=-\frac{ D_{{\bf U}_2}(-k)}{ D_{{\bf U}_2}(k)}\;.
\end{equation}
Introducing the notation $\rho_{R,L}\equiv  D_{{\bf U}_2}(k) r_{R,L}$ and $\tau_{R,L}= D_{{\bf U}_2}(k) t_{R,L}$ allows us to rewrite \eqref{1f} as
\begin{eqnarray}
F_{\lambda}(k,a;S,{\bf U}_1)&=&\frac{1}{D_{{\bf U}_2}(k)}\left\{ D_{{\bf U}_2}(k)e^{-ikL}\left[ C_{{\bf U}_1}^++k^2C_{{\bf U}_1}^--2k(1-\det({\bf U}_1))
\right]\right.\nonumber\\
&&+D_{{\bf U}_2}(-k)e^{ikL}\left[ C_{{\bf U}_1}^++k^2C_{{\bf U}_1}^-+2k(1-\det({\bf U}_1))\right]\nonumber\\
&&+(\rho_Re^{ik(2a-L)}+\rho_Le^{-ik(2a-L)})\left[C_{{\bf U}_1}^+-k^2C_{{\bf U}_1}^-\right]\nonumber\\
&&\left.+2k(\rho_Re^{ik(2a-L)}-\rho_Le^{-ik(2a-L)})(u_{11}-u_{22})+4k(u_{21}\tau_R+u_{12}\tau_L)\right\}\;.\;\;\;\;\;\;\;\label{2f}
\end{eqnarray}

By exploiting now Euler's parametrization of the group $U(2)$ for ${\bf U}_1$, that is
\begin{equation}\label{24}
{\bf U}_1=e^{i \alpha}\left[ \mathbb{I} \cos (\beta )+i \sin (\beta ) \left(n_1 \sigma _1+n_2
   \sigma _2+n_3 \sigma _3\right)\right]\;,
\end{equation}
with $\alpha\in [-\pi,\pi]$ and  $\beta\in [-\pi/2,\pi/2]$,
one finds the relations
\begin{eqnarray}\label{25}
&&C_{{\bf U}_1}^{\mp}=1+\det({\bf U}_1)\pm{\rm tr}({\bf U}_1)=2e^{i\alpha}(\cos(\alpha)\pm\cos(\beta)),\\
&&1-\det({\bf U}_1)=-2 ie^{i\alpha}\sin(\alpha),\quad u_{11}-u_{22}=2 i n_3e^{i\alpha}\sin(\beta)\\
&&u_{12}=i e^{i \alpha}\sin(\beta)(n_1-in_2), \quad u_{21}=i e^{i \alpha}\sin(\beta)(n_1+in_2),
\end{eqnarray}
which can be used in (\ref{2f}) to obtain the following expression
\begin{eqnarray}\label{26}
F_{\lambda}(k,a;S,{\bf U}_1)&=&\frac{2 e^{i\alpha}}{D_{{\bf U}_2}(k)}\big\{ D_{{\bf U}_2}(k)e^{-ikL}\left[ \cos(\alpha)-\cos(\beta)+k^2( \cos(\alpha)+\cos(\beta))
+2ik\sin(\alpha)\right]\nonumber\\
&&+D_{{\bf U}_2}(-k)e^{ikL}\left[ \cos(\alpha)-\cos(\beta)+k^2( \cos(\alpha)+\cos(\beta))-2ik\sin(\alpha)\right]\nonumber\\
&&+(\rho_Re^{ik(2a-L)}+\rho_Le^{-ik(2a-L)})\left[\cos(\alpha)-\cos(\beta)-k^2( \cos(\alpha)+\cos(\beta))\right]\nonumber\\
&&+2i kn_3\sin(\beta)(\rho_Re^{ik(2a-L)}-\rho_L e^{-ik(2a-L)})\nonumber \\
&&+2ik\sin(\beta)((n_1+in_2)\tau_R+(n_1-in_2)\tau_L)\big\}\;.
\end{eqnarray}
One final remark regards the overall factor in (\ref{26}). It is easy to realize, from the definition in (\ref{Usec}), that $D_{{\bf U}_2}(k)$ has no poles.
This implies that the factor $2 e^{i\alpha}(D_{{\bf U}_2}(k))^{-1}$ does not contribute to the zeroes of the function $F_{\lambda}(k,a;S,{\bf U}_1)$ and can, hence, be
safely discarded. We can therefore conclude that the function $F_{\lambda}(k,a;S,{\bf U}_1)$ has the same zeroes as the following function
\begin{equation}\label{32}
h_{\lambda}(k,a;S,{\bf U}_1)=\frac{e^{-i\alpha}}{2}D_{{\bf U}_2}(k)F_{\lambda}(k,a;S,{\bf U}_1)\;,
\end{equation}
which is the one we will utilize in order to analyze the spectral zeta function of our piston configuration.

\section{The spectral zeta function and Casimir energy}

The function $h_{\lambda}(k,a;S,{\bf U}_1)$ can be used to derive an expression for the spectral zeta function
associated with the piston configuration which is defined in terms of the eigenvalues $\alpha$ of our system
as follows
\begin{equation}\label{27}
\zeta(s,a)=\sum_{\alpha>0}\alpha^{-2s}\;.
\end{equation}
The above zeta function is known to be convergent for $\Re(s)>D/2$ \cite{elizalde95,elizalde94,kirsten01} and can be analytically continued
to a meromorphic function in the whole complex plane possessing only simple poles. The spectral zeta function can be utilized to compute the Casimir energy
of suitable quantum systems \cite{bordag09,bordag01,bytse03,elizalde95,elizalde94,kirsten01}, and in particular for the piston configuration under consideration in this work. In this framework, as outlined in Section \ref{sec2new}, the Casimir energy of a piston is expressed as
(replacing $\zeta_{\widehat K} (s)$ with $\zeta (s,a)$)
\begin{equation}\label{29}
E_{Cas} (a) =\frac{1}{2}\left[ \textrm{FP}\,\zeta\left(-\frac{1}{2},a\right)+\ln\mu^{2}\textrm{Res}\,\zeta\left(-\frac{1}{2},a\right)\right]\;.
\end{equation}         
From the expression for the Casimir energy in (\ref{29}) one obtains the Casimir force acting on the piston by simply differentiating with respect to the
position $a$ of the piston, that is
\begin{equation}\label{30}
F_{\textrm{Cas}}(a)=-\frac{\partial}{\partial a}E_{\textrm{Cas}}(a)\;.
\end{equation}
From the formulas (\ref{29}) and (\ref{30}) it is not very difficult to realize that
the Casimir force acting on the piston is a uniquely defined quantity only if the residue of the spectral zeta function
at $s=-1/2$ is independent of the position $a$ of the piston, because this ensures that the result obtained is independent of the regularization parameter $\mu$.

The eigenvalues $\alpha$ of our system are only known implicitly as the positive zeroes of the function $h_{\lambda}(k,a;S,{\bf U}_1)$
through the relation $\alpha^2=k^2+\lambda^2$.
One can, therefore, employ a contour integral representation, based on Mittag-Leffler's theorem, to write the spectral zeta function as follows \cite{bordag96,bordag96b,kirsten01}
\begin{equation}\label{31}
\zeta(s,a)=\frac{1}{2\pi i}\sum_{\lambda}d(\lambda)\int_{\gamma}\left(k^{2}+\lambda^{2}\right)^{-s}\frac{\partial}{\partial k}\ln h_{\lambda}(k,a;S,{\bf U}_1)\,\diff k\;,
\end{equation}
valid in the region of the complex plane $\Re(s)>D/2$. Here, $\gamma$ represents a contour that encloses, in the counterclockwise direction, all
the positive zeroes of the function $h_{\lambda}(k,a;S,{\bf U}_1)$. In addition, $d(\lambda)$ denotes the degeneracy of the
eigenvalues $\lambda$ of the Laplacian on the transverse manifold $N$. In order to analyze the Casimir energy of the system and the corresponding force,
the expression in (\ref{31}) for $\zeta(s,a)$ needs to be analytically extended to a neighborhood of the point $s=-1/2$.
The first step in the analytic continuation consists of deforming
the contour $\gamma$ to the imaginary axis \cite{kirsten01}. Before performing the contour deformation, it is very important to analyze
the small-$k$ behavior of the function $h_{\lambda}(k,a;S,{\bf U}_1)$. By using the definition (\ref{32}) and the expression (\ref{26})
one obtains the following asymptotic behavior as $k\to 0$
\begin{eqnarray}
h_{\lambda}(k,a;S,{\bf U}_1)&=&\big\{8\cos(\theta)\cos(\alpha)-8\cos(\gamma)\cos(\beta)+4\sin(\theta)[L(\cos(\alpha)-\cos(\beta))-2\sin(\alpha)]\nonumber\\
&+&4(\cos(\gamma)-\cos(\theta))[a(a-L)(\cos(\alpha)-\cos(\beta))+L\sin(\alpha)]\nonumber\\
&-&4(2a-L)[(\cos(\alpha)-\cos(\beta))q_{3}\sin(\gamma)+(\cos(\gamma)-\cos(\theta))n_{3}\sin(\beta)]\nonumber\\
&+&8\sin(\beta)\sin(\gamma)(n_{1}q_{1}+n_{2}q_{2}-n_{3}q_{3})\big\}k^{2}+O(k^{4})\;.
\end{eqnarray}
Since $h_{\lambda}(k,a;S,{\bf U}_1)$ is of order $k^{2}$ as $k\to 0$, a simple contour deformation to the imaginary axis would
allow the integral to acquire an unwanted contribution from the origin $k=0$. In order to avoid this spurious contribution we replace the
representation (\ref{31}) of the spectral zeta function with the following one
\begin{equation}\label{34}
\zeta(s,a)=\frac{1}{2\pi i}\sum_{\lambda}d(\lambda)\int_{\gamma}\left(k^{2}+\lambda^{2}\right)^{-s}\frac{\partial}{\partial k}\ln\left[\frac{ h_{\lambda}(k,a;S,{\bf U}_1)}{k^{2}}\right]\,\diff k\;.
\end{equation}

By exploiting the fact that the function $h_{\lambda}(k,a;S,{\bf U}_1)$ satisfies the property
\begin{equation}\label{35}
h_{\lambda}(ik,a;S,{\bf U}_1)=h_{\lambda}(-ik,a;S,{\bf U}_1)\;,
\end{equation}
which can be proved by noticing that for any $w\in\mathbb{C}$, $\rho_R(-w)=\rho_L(w)$ and $\tau_R(-w)=\tau_L(w)$, the contour deformation to the imaginary axis leads to the expression
\begin{equation}\label{36}
\zeta(s,a)=\sum_{\lambda}d(\lambda)\zeta_{\lambda}(s,a)\;,
\end{equation}
where we have introduced the zeta function
\begin{equation}\label{37}
\zeta_{\lambda}(s,a)=\frac{\sin(\pi s)}{\pi}\int_{\lambda}^{\infty}\left(z^{2}-\lambda^{2}\right)^{-s}\frac{\partial}{\partial z}\ln\left[\frac{ h_{\lambda}(iz,a;S,{\bf U}_1)}{ z^{2}}\right]\diff z\;.
\end{equation}
The integral representation (\ref{37}) is valid in the region of the complex plane $1/2<\Re(s)<1$. The upper bound on the region of validity
is obtained by requiring the integral to be convergent at the lower limit of integration and by noticing that, as $z\to \lambda$, the integrand behaves as
\begin{equation}\label{38}
(z^{2}-\lambda^{2})^{-s}\frac{\partial}{\partial z}\ln\left[\frac{ h_{\lambda}(i z,a;S,{\bf U}_1)}{z^{2}}\right]\sim (z-\lambda)^{-s}\;.
\end{equation}
As $z\to\infty$ the function $h_{\lambda}(i z,a;S,{\bf U}_1)$ displays, instead, the following behavior
\begin{eqnarray}\label{39}
\lefteqn{h_{\lambda}(i z,a;S,{\bf U}_1)=D_{{\bf U}_2}(i z)e^{ z L}\Big[\cos(\alpha)-\cos(\beta)}\nonumber\\
&&-z^{2}( \cos(\alpha)+\cos(\beta))-2z\sin(\alpha)\Big][1+\varepsilon(iz,a)]\;,
\end{eqnarray}
where $\varepsilon(iz,a)$ represents exponentially small terms. The expression (\ref{39}) allows us to conclude that, as $z\to\infty$, the integrand
in (\ref{37}) behaves as
\begin{equation}\label{40}
(z^{2}-\lambda^{2})^{-s}\frac{\partial}{\partial z}\ln\left[\frac{ h_{\lambda}(i z,a;S,{\bf U}_1)}{z^{2}}\right]\sim  Lz^{-2s}\;,
\end{equation}
which together with the requirement that the integral representation (\ref{37}) be convergent at the upper limit of integration, provides the lower bound
$\Re(s)>1/2$.

In order to analyze the Casimir energy and the corresponding force, we need to extend the definition of the zeta function in (\ref{37}) to the region
of the complex plane $\Re(s)\leq 1/2$. This is accomplished by simply subtracting and then adding in the integral representation (\ref{37})
a suitable number of terms of the asymptotic expansion as $z\to\infty$ of $\ln\left[z^{-2}h_{\lambda}(iz,a;S,{\bf U}_1)\right]$.
By using the definition (\ref{Usec}) we can write a formula which we can use as a starting point of the asymptotic expansion
\begin{eqnarray}\label{41}
\ln\left[z^{-2}h_{\lambda}(iz,a;S,{\bf U}_1)\right]&\simeq&zL-2\ln z+\ln\Psi(z;\theta,\gamma)+\ln\Psi(z;\alpha,\beta)\;,
\end{eqnarray}
where we have discarded the exponentially small terms and we have introduced, for convenience, the function
\begin{equation}\label{41a}
\Psi(z;x,y)=m_{-}(x,y)-2z\sin x-z^{2}\,m_{+}(x,y)\;,
\end{equation}
with
\begin{equation}\label{42}
m_{\pm}(x,y)=\cos x\pm\cos y\;.
\end{equation}
From the expressions (\ref{41})-(\ref{42}) it is not difficult to see that the specific form of the asymptotic expansion depends on whether or not
the coefficients $m_{+}(\alpha,\beta)$, $\sin\alpha$, $m_{+}(\theta,\gamma)$, and $\sin\theta$ vanish. In order to consider all the cases simultaneously
we introduce the function
\begin{equation}
\delta(x)=\left\{\begin{array}{ll}
1 & \textrm{if}\; x=0\\
0 & \textrm{if}\; x\neq 0
\end{array}\right.\;,
\end{equation}
and rewrite the logarithm of (\ref{41a}) as follows
\begin{eqnarray}\label{43}
\ln\Psi(z;x,y)&=&\left[2-\delta(m_{+}(x,y))(1+\delta(\sin x))\right]\ln z+\tau(x,y)\nonumber\\
&+&\left[1-\delta(m_{+}(x,y))\right]\ln\left[1+\frac{2\sin x}{m_{+}(x,y)z}-\frac{m_{-}(x,y)}{m_{+}(x,y)z^{2}}\right]\nonumber\\
&+&\delta(m_{+}(x,y))[1-\delta(\sin x)]\ln\left[1-\frac{m_{-}(x,y)}{2 z \sin x}\right]\; ,
\end{eqnarray}
where
\begin{eqnarray}
\tau(x,y)&=&\left[1-\delta(m_{+}(x,y))\right]m_{+}(x,y)+\delta(m_{+}(x,y))[1-\delta(\sin x)]\ln(2\sin x)\nonumber\\
&+&\delta(m_{+}(x,y))\delta(\sin x)\ln[m_{-}(x,y)]\;.
\end{eqnarray}
The large-$z$ asymptotic expansion of the quantity in (\ref{43}) can be obtained by following the argument presented in \cite{mukibo-lmp15}.
More explicitly one finds
\begin{eqnarray}\label{44}
\ln\Psi(z;x,y)&=&\left[2-\delta(m_{+}(x,y))(1+\delta(\sin x))\right]\ln z+\tau(x,y)+\sum_{n=1}^{\infty}\frac{\omega_{n}(x,y)}{z^{n}}\;,
\end{eqnarray}
where (cf. \cite{mukibo-lmp15})
\begin{eqnarray}\label{45}
\omega_{n}(x,y)&=&\left[1-\delta(m_{+}(x,y))\right](-1)^{n+1}\sum_{n=0}^{\left[\frac{n}{2}\right]}\frac{2^{n-2j}\Gamma(n-j)}{j!\Gamma(n-2j+1)}(\sin x)^{n-2j}\frac{m^{j}_{-}(x,y)}{m_{+}^{n-j}(x,y)}\nonumber\\
&-&\delta(m_{+}(x,y))[1-\delta(\sin x)]\frac{(\cot x)^{n}}{n}\;.
\end{eqnarray}

By exploiting the formula (\ref{44}) it is not very difficult to write the large-$z$ asymptotic expansion of (\ref{41}), that is
\begin{eqnarray}\label{46}
\ln\left[z^{-2}h_{\lambda}(iz,a;S,{\bf U}_1)\right]&\simeq&zL+\chi(\theta,\gamma,\alpha,\beta)\ln z+\tau(\theta,\gamma)+\tau(\alpha,\beta)\nonumber\\
&+&\sum_{n=1}^{\infty}\frac{\omega_{n}(\theta,\gamma)+\omega_{n}(\alpha,\beta)}{z^{n}}\;,
\end{eqnarray}
where we have introduced the function
\begin{equation}\label{46a}
\chi(\theta,\gamma,\alpha,\beta)=2-\delta(m_{+}(\theta,\gamma))[1+\delta(\sin\theta)]
-\delta(m_{+}(\alpha,\beta))[1+\delta(\sin\alpha)]\;.
\end{equation}
The above asymptotic expansion can now be used to perform the analytic continuation of the spectral zeta function. By subtracting and then adding
the first $N$ terms of the asymptotic expansion (\ref{46}) in the integrand of (\ref{37}) we get
\begin{equation}\label{47}
\zeta(s,a)=Z(s,a)+\sum_{i=-1}^{N}A_{i}(s)\;,
\end{equation}
where $Z(s,a)$ is an analytic function in the region $\Re(s)>(d-N-1)/2$ and has the form
\begin{eqnarray}\label{48}
Z(s,a)&=&\frac{\sin(\pi s)}{\pi}\sum_{\lambda}d(\lambda)\int_{\lambda}^{\infty}\left(z^{2}-\lambda^{2}\right)^{-s}\frac{\partial}{\partial z}\Bigg\{\ln{\left[ \frac{ h_{\lambda}(iz,a;S,{\bf U}_1)}{ z^{2}}\right]}-zL\nonumber\\
&-&\chi(\theta,\gamma,\alpha,\beta)\ln z-\tau(\theta,\gamma)-\tau(\alpha,\beta)-\sum_{n=1}^{N}\frac{\omega_{n}(\theta,\gamma)+\omega_{n}(\alpha,\beta)}{z^{n}}\Bigg\}\diff z\;.\label{big-z}
\end{eqnarray}
The remaining quantities in (\ref{47}), i.e. $A_{i}(s)$, are obtained by integrating the terms of the asymptotic asymptotic expansion that have been added back and
are meromorphic functions of $s$ possessing only isolated simple poles. It is not difficult to prove that
\begin{eqnarray}\label{49}
A_{-1}(s)&=&\frac{L}{2\sqrt{\pi}\Gamma(s)}\Gamma\left(s-\frac{1}{2}\right)\zeta_{N}\left(s-\frac{1}{2}\right)\;,\\
A_{0}(s)&=&\frac{1}{2}\chi(\theta,\gamma,\alpha,\beta)\zeta_{N}(s)\;,\;\;
\end{eqnarray}
and, for $i\geq1$,
\begin{equation}\label{50}
A_{i}(s)=-\frac{\omega_{i}(\theta,\gamma)+\omega_{i}(\alpha,\beta)}{\Gamma(s)\Gamma\left(\frac{i}{2}\right)}\Gamma\left(s+\frac{i}{2}\right)\zeta_{N}\left(s+\frac{i}{2}\right)\;,
\end{equation}
where in the previous expressions we have used the following definition of the spectral zeta function associated with the Laplacian $-\Delta_{N}$ on the manifold $N$
\begin{equation}
\zeta_{N}(s)=\sum_{\lambda}d(\lambda)\lambda^{-2s}\;.
\end{equation}
Before exploiting these results for the Casimir energy, let us remark that the above equations are also perfectly suited to compute the heat kernel coefficients for the piston setting.
It is known that only the $A_j (s)$, $j=-1,0,1,...$, contribute to the coefficients and (\ref{46}) and (\ref{46a}) clearly show how contributions split into $(\theta , \gamma )$ and $(\alpha , \beta )$
dependent parts, which have been treated in \cite{mukibo-lmp15}. Results for heat kernel coefficients will therefore simply be sums of results given in \cite{mukibo-lmp15} and we will not present more details in this context.

We will now employ the analytically continued expression of the spectral zeta function in (\ref{47}) and the definition in (\ref{29})
to derive a formula for the Casimir energy of the piston.
By setting $N=D$ in (\ref{47}) we obtain a representation for the spectral zeta function valid in the region
$-1<\Re(s)<1$ and, hence, suitable for the calculation of the Casimir energy. According to the definition (\ref{29}) the Casimir energy is computed by setting
$s=\epsilon-1/2$ in (\ref{47}) and by subsequently taking the limit $\epsilon\to0$. During this limiting process, the meromorphic structure of the spectral zeta
function $\zeta_{N}(s)$ plays an important role. In accordance with the general theory of spectral zeta functions, \cite{gilkey,kirsten01} one has
\begin{eqnarray}\label{51}
\zeta_{N}(\epsilon -n)&=&\zeta_{N}(-n)+\epsilon\zeta^{\prime}_{N}(-n)+O(\epsilon^{2})\;,\\
\zeta_{N}\left(\epsilon+\frac{d-k}{2}\right)&=&\frac{1}{\epsilon}\textrm{Res}\,\zeta_{N}\left(\frac{d-k}{2}\right)+\textrm{FP}\,\zeta_{N}\left(\frac{d-k}{2}\right)+O(\epsilon)\;,\\
\zeta_{N}\left(\epsilon-\frac{2n+1}{2}\right)&=&\frac{1}{\epsilon}\textrm{Res}\,\zeta_{N}\left(-\frac{2n+1}{2}\right)+\textrm{FP}\,\zeta_{N}\left(-\frac{2n+1}{2}\right)+O(\epsilon)\;,
\end{eqnarray}
where $n\in\mathbb{N}_{0}$ and $k=\{0,\ldots,d-1\}$.
Since $Z(s,a)$ is an analytic function for $-1<\Re(s)<1$, we can simply set $s=-1/2$ in its expression. For the terms $A_{i}(s)$ we find instead (c.f. \cite{fucci-npb15})
\begin{equation}\label{52}
A_{-1}\left(\epsilon-\frac{1}{2}\right)=\frac{L\,\zeta_{N}(-1)}{4\pi\varepsilon}
  +\frac{L}{4\pi}\left[\zeta'_{N}(-1)+(2\ln 2-1)\zeta_{N}(-1)\right]+O(\varepsilon)\;,
\end{equation}
\begin{eqnarray}\label{53}
A_{0}\left(\epsilon-\frac{1}{2}\right)&=&\frac{1}{2}\chi(\theta,\gamma,\alpha,\beta)
\left[\frac{1}{\epsilon}\textrm{Res}\,\zeta_{N}\left(-\frac{1}{2}\right)+\textrm{FP}\,\zeta_{N}\left(-\frac{1}{2}\right)\right]+O(\varepsilon)\;,
\end{eqnarray}
and
\begin{eqnarray}\label{54}
\lefteqn{\sum_{i=1}^{D}A_{i}\left(\epsilon-\frac{1}{2}\right)=\frac{1}{\epsilon}\Bigg[\frac{\omega_{1}(\theta,\gamma)+\omega_{1}(\alpha,\beta)}{2\pi}\zeta_{N}(0)+\sum_{i=2}^{D}
\frac{\omega_{i}(\theta,\gamma)+\omega_{i}(\alpha,\beta)}{2\sqrt{\pi}\Gamma\left(\frac{i}{2}\right)}\Gamma\left(\frac{i-1}{2}\right)}\nonumber\\
&&\times\textrm{Res}\,\zeta_{N}\left(\frac{i-1}{2}\right)\Bigg]+\frac{\omega_{1}(\theta,\gamma)+\omega_{1}(\alpha,\beta)}{2\pi}\left[\zeta'_{N}(0)+2(\ln 2-1)\zeta_{N}(0)\right]\nonumber\\
&&+\sum_{i=2}^{D}\frac{\omega_{i}(\theta,\gamma)+\omega_{i}(\alpha,\beta)}{2\sqrt{\pi}\Gamma\left(\frac{i}{2}\right)}\Gamma\left(\frac{i-1}{2}\right)\Bigg[\textrm{FP}\,\zeta_{N}\left(\frac{i-1}{2}\right)+
\left(2-\gamma-2\ln 2+\Psi\left(\frac{i-1}{2}\right)\right)\nonumber\\
&&\times\textrm{Res}\,\zeta_{N}\left(\frac{i-1}{2}\right)\Bigg]+O(\varepsilon)\;.
\end{eqnarray}
The above results together with the formula (\ref{29}) allow us to write an explicit expression for the Casimir energy of the piston configuration as follows
\begin{eqnarray}\label{55}
\lefteqn{E_{\textrm{Cas}}(a)=\frac{1}{2}\left(\frac{1}{\varepsilon}+\ln\mu^{2}\right)\Bigg[\frac{L}{4\pi}\zeta_{N}(-1)+\frac{1}{2}\chi(\theta,\gamma,\alpha,\beta)\textrm{Res}\,\zeta_{N}\left(-\frac{1}{2}\right)+\frac{\omega_{1}(\theta,\gamma)+\omega_{1}(\alpha,\beta)}{2\pi}\zeta_{N}(0)}\nonumber\\
&+&\sum_{i=2}^{D}
\frac{\omega_{i}(\theta,\gamma)+\omega_{i}(\alpha,\beta)}{2\sqrt{\pi}\Gamma\left(\frac{i}{2}\right)}\Gamma\left(\frac{i-1}{2}\right)\textrm{Res}\,\zeta_{N}\left(\frac{i-1}{2}\right)\Bigg]+\frac{1}{2}Z\left(-\frac{1}{2},a\right)\nonumber\\
&+&\frac{L}{8\pi}\left[\zeta'_{N}(-1)+(2\ln 2-1)\zeta_{N}(-1)\right]+\frac{1}{4}\chi(\theta,\gamma,\alpha,\beta)\textrm{FP}\,\zeta_{N}\left(-\frac{1}{2}\right)\nonumber\\
&+&\frac{\omega_{1}(\theta,\gamma)+\omega_{1}(\alpha,\beta)}{2\pi}\left[\zeta'_{N}(0)+2(\ln 2-1)\zeta_{N}(0)\right]+\sum_{i=2}^{D}\frac{\omega_{i}(\theta,\gamma)+\omega_{i}(\alpha,\beta)}{2\sqrt{\pi}\Gamma\left(\frac{i}{2}\right)}\Gamma\left(\frac{i-1}{2}\right)\nonumber\\
&\times&\Bigg[\textrm{FP}\,\zeta_{N}\left(\frac{i-1}{2}\right)+
\left(2-\gamma-2\ln 2+\Psi\left(\frac{i-1}{2}\right)\right)\textrm{Res}\,\zeta_{N}\left(\frac{i-1}{2}\right)\Bigg]+O(\varepsilon)\;.
\end{eqnarray}
The above expression clearly shows that the Casimir energy of the piston configuration is, in general, not a well-defined quantity.
The ambiguity in the force is proportional to $\zeta_{N}(-1)$, $\zeta_{N}(0)$, and the $\textrm{Res}\,\zeta_{N}\left((i-1)/2\right)$ with $i=0,\ldots,D$. These quantities
depend, in turn, only on the geometry of the transverse manifold $N$ and the boundary conditions imposed on the fields propagating on $N$ through the coefficients $a^{N}_{k/2}$ of the asymptotic expansion of the heat kernel associated with $\Delta_{N}$. This is due to the well-known relations with $n\in\mathbb{N}_{0}$ \cite{gilkey,kirsten01}
\begin{eqnarray}\label{56}
  \Gamma\left(\frac{d-k}{2}\right)\textrm{Res}\,\zeta_{N}\left(\frac{d-k}{2}\right)&=&a^{N}_{\frac{k}{2}}\;,\nonumber\\
  \Gamma\left(-\frac{2n+1}{2}\right)\textrm{Res}\,\zeta_{N}\left(-\frac{2n+1}{2}\right)&=&a^{N}_{\frac{d+2n+1}{2}}\;,\nonumber\\
  \frac{(-1)^{n}}{\Gamma(n+1)}\zeta_{N}(-n)&=&a^{N}_{\frac{d}{2}+n}\;.
\end{eqnarray}
While the Casimir energy is generally ambiguous, the Casimir force acting on the piston is a well-defined quantity since the terms responsible
for the ambiguity in the energy do not depend on the position of the piston. In fact, by using (\ref{55}) and the definition provided in (\ref{30})
we obtain the following simple expression for the Casimir force acting on the piston
\begin{equation}\label{57}
F_{\textrm{Cas}}(a)=-\frac{1}{2}\frac{d}{da}Z\left(-\frac{1}{2},a\right)=\frac{1}{2\pi}\sum_{\lambda}d(\lambda)\frac{d}{da}J_\lambda(a).
\end{equation}
where $Z\left(-\frac{1}{2},a\right)$ is given by formula \eqref{big-z}, and we have introduced the notation
\begin{equation}
J_\lambda(a)\equiv\int_{\lambda}^{\infty}\left(z^{2}-\lambda^{2}\right)^{\frac{1}{2}}\partial_z\left[\ln( h_{\lambda}(iz,a;S,{\bf U}_1)-As(z; S,{\bf U}_1)\right]\diff z   \:,
\end{equation}
with $As(z; S,{\bf U}_1)$ being the asymptotic terms subtracted in equation \eqref{big-z}; note, that these terms do not depend on the position of the piston $a$. If we integrate by parts in $J_\lambda(a)$, and take into account that the boundary terms cancel, we can write
\begin{equation}
J_\lambda(a)=-\int_{\lambda}^{\infty}\frac{z}{\left(z^{2}-\lambda^{2}\right)^{\frac{1}{2}}}\left[\ln( h_{\lambda}(iz,a;S,{\bf U}_1)-As(z; S,{\bf U}_1)\right]\diff z.
\end{equation}
Hence the Casimit force can finally be written as
\begin{equation}\label{57-bis}
F_{\textrm{Cas}}(a)=-\frac{1}{2\pi}\sum_{\lambda}d(\lambda)\int_{0}^{\infty}\partial_a\left[\ln( h_{\lambda}(i\sqrt{w^2+\lambda^2},a;S,{\bf U}_1)\right]\diff w,
\end{equation}
after performing the change of variables $w=\sqrt{z^2-\lambda^2}$. The formula \eqref{57-bis} for the Casimir force will be used in the next section to generate graphs of the Casimir force on the piston for different geometries and boundary conditions.


\section{Casimir force for particular piston geometries}

It is clear from the expression \eqref{57-bis} that the Casimir force acting on the piston can be obtained numerically once the manifold $N$ and the boundary
conditions have been specified. In this section we consider the following two manifolds $N$: the two-dimensional disk and the $d$-dimensional sphere.
Before proceeding with these two cases we would like to make a remark about the piston configuration constructed from a generalized torus.
This particular piston configuration is obtained by imposing periodic boundary conditions at $x=0$ and $x=L$. In this case the left edge and the right one of the piston
configuration are identified. Periodic boundary conditions can be obtained by setting $\alpha=\pi/2$, $\beta=\pm\pi/2$, and $n_{1}=\mp 1$ in ${\bf U}_{1}$ \cite{mukibo-lmp15}.
With this particular choice of parameters, it is not difficult to realize that the terms with the dependence on the position of the piston $a$ in
$h_{\lambda}(k,a;S,{\bf U}_1)$ in (\ref{32}) vanish identically. This implies that in a generalized torus the piston itself does not incur any force.
This result should be expected because identifying the two edges of the piston is equivalent to reducing the piston configuration to a single chamber. More generally, for any configuration where $\alpha=\pi/2,3\pi/2$, $\beta=\pm\pi/2$, and $n_3=0$ all the terms dependent on the position of the piston $a$ that appear in $h_{\lambda}(k,a;S,{\bf U}_1)$ (see equation\eqref{32}) vanish identically. We can, therefore, conclude that in these situations there is \emph{no} Casimir force acting on the piston. In addition, if the selfa-djoint extension that characterises the piston gives rise to an opaque piston wall, i.e. $r_R=r_L=0$ the Casimir force vanishes as well since all the $a$-dependent terms in $h_{\lambda}(k,a;S,{\bf U}_1)$ are proportional to either $\rho_R$ or $\rho_L$. Nevertheless, the special case of $\alpha=\pi/2=-\beta$ and $n_3=0\Rightarrow n_1=\cos(\xi),\, n_2=\sin(\xi)$ is of great interest when the cross section of the piston geometry degenerates to a point. In this case we interpret the free parameter $\xi$ as the quasi-momentum of a one-dimensional crystal lattice where the lattice points are mimicked by identical point-supported potentials, generalising the result of reference \cite{bor-fphys19}. For the examples that we consider in this section we will assume, for simplicity, that $L=1$.

\subsection{The $d$-dimensional sphere}

\begin{figure}[htbp]
	\begin{center}
		\includegraphics[width=0.3\linewidth]{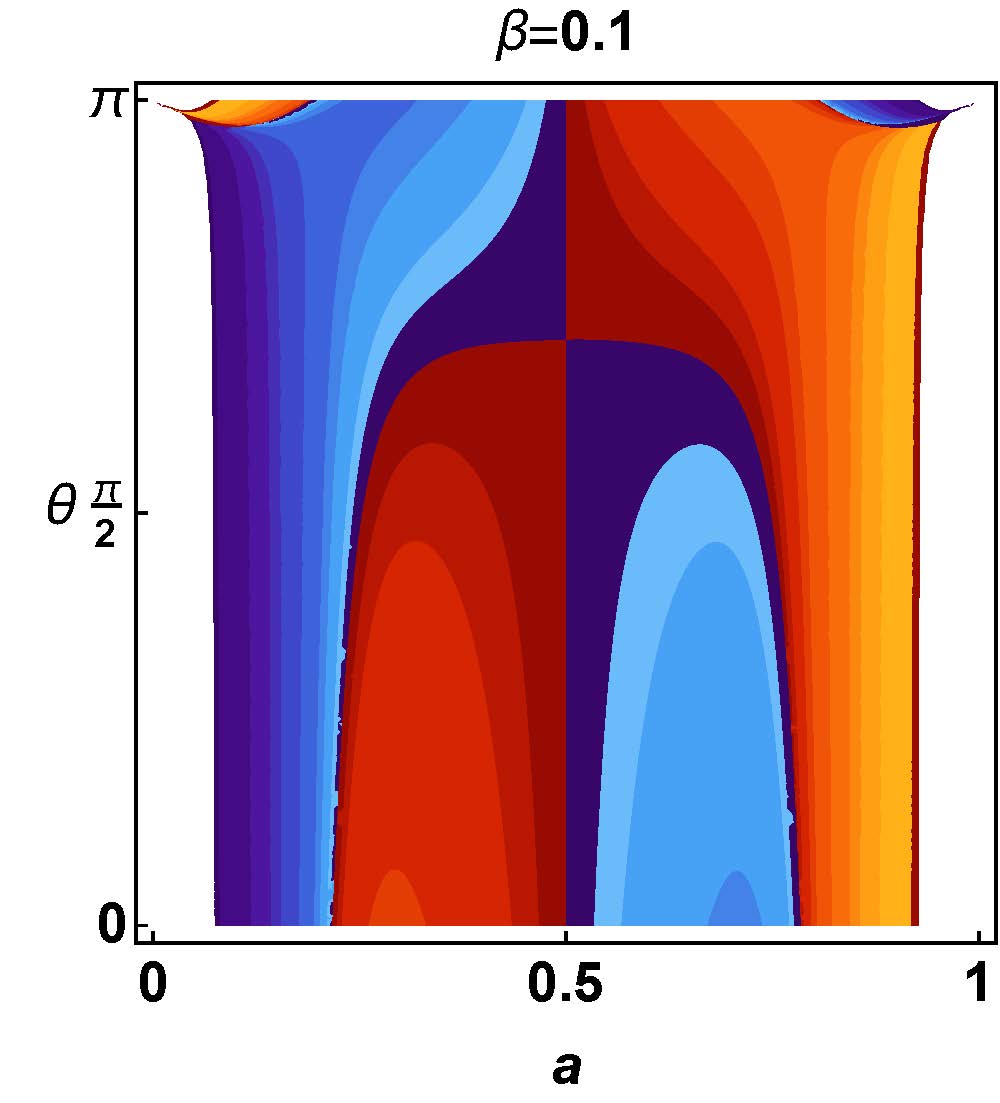}\quad \includegraphics[width=0.3\linewidth]{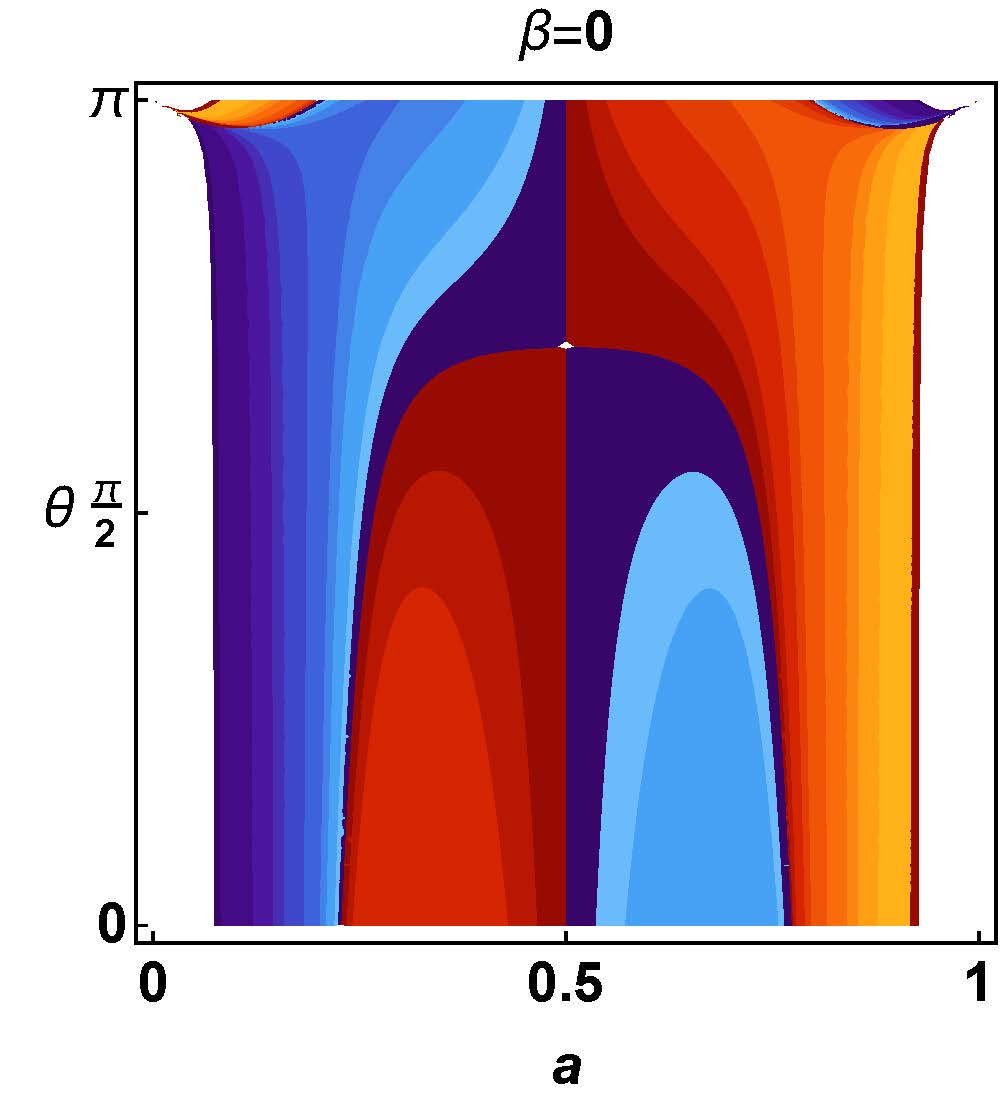} \quad \includegraphics[width=0.3\linewidth]{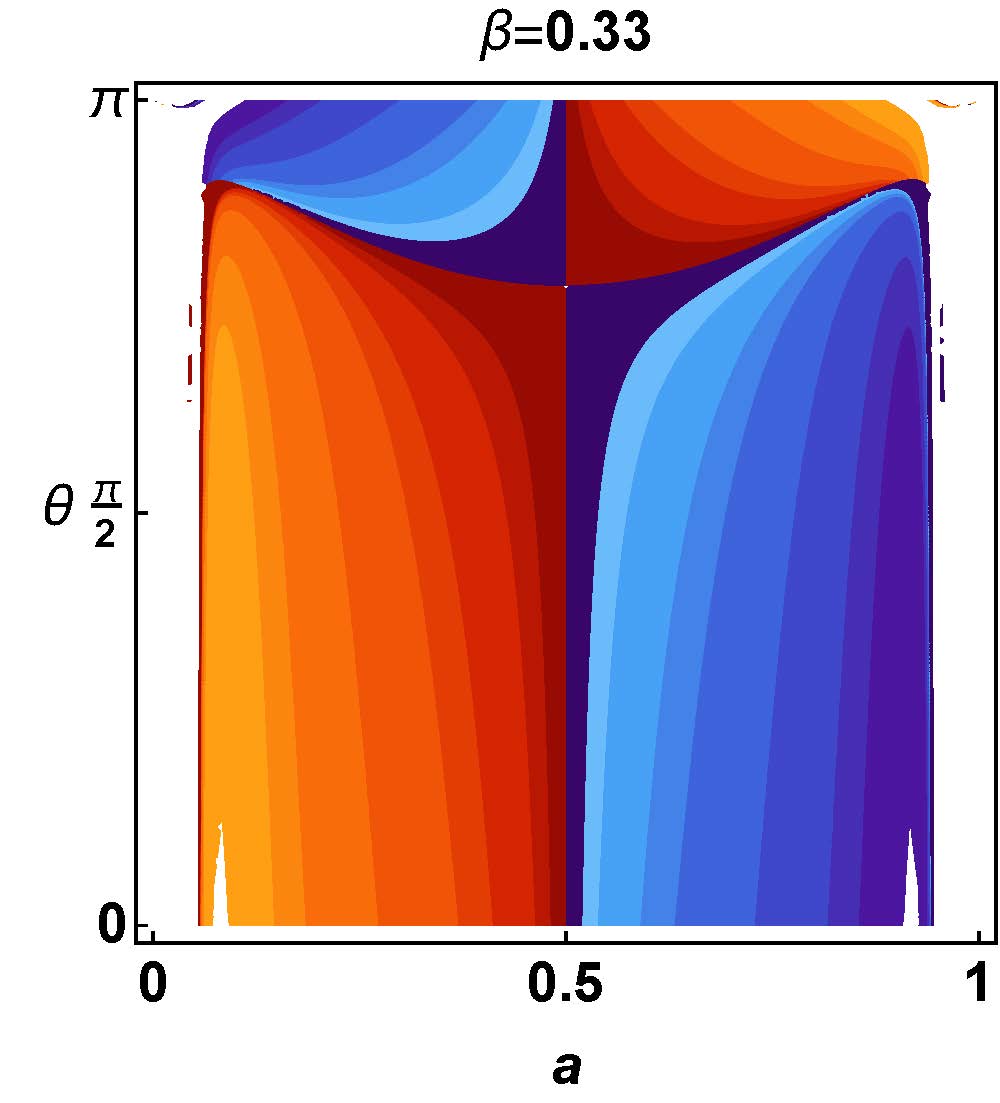}
		\caption{(color online) Behavior of the Casimir force \eqref{57-bis} as a function of the parameter $\theta$ of the piston characterised by ${\bf U}_2$ and the position $a$ of the piston, for different values of $\beta$. The rest of the parameters are fixed to $L=1$, $\alpha=2.8$, $n_1=q_1=1$, and $\gamma=0$. The curves separating positive force (red color scale) and negative force (blue color scale) correspond to zero Casimir force situations.}
		\label{fig-2}
	\end{center}
\end{figure}

In this example we consider the base manifold to be a $d$-dimensional sphere. The eigenvalues of the Laplacian $\Delta_{N}$ on a $d$-dimensional sphere are known to be
\begin{equation}\label{59}
\lambda^2=l(l+d-1)\;,
\end{equation}
with $l\in\mathbb{N}_{0}$, and the associated degeneracy has the form
\begin{equation}\label{60}
d(\nu)=(2l+d-1)\frac{(l+d-2)!}{l!(d-1)!}\;.
\end{equation}
In order to obtain specific graphs of the Casimir force on the piston as a function of the position $a$ we set $d=2$ and we use the eigenvalues and degeneracy (\ref{59})
and (\ref{60}) in the expression (\ref{57}). Once particular boundary conditions are chosen, a numerical analysis of the Casimir force \eqref{57-bis} can be performed. It is important to point out that the dimension $d=2$ has been chosen only for simplicity and that our formula for the Casimir force (\ref{57}) holds for any dimension $d$. Figures \ref{fig-2}-\ref{fig-5} show the behavior of the Casimir force on the piston for specific boundary conditions imposed on the field.

The figures have been generated by utilizing a two colors scheme. The blue and red areas denote those regions in the space of parameters in which the Casimir force is negative, respectively, positive. The shade of the color gives a measure of the magnitude of the force on the piston: The darker the color, the smaller the magnitude, the lighter the color the higher the magnitude. The white areas appearing in graphs are those in which the magnitude of the force exceeds the range of the graph. However, the white areas at the two edges of the piston, $x=0$ and $x=1$, reflect the fact that the Casimir force grows without bounds as one approaches the edges. The growth is positive (negative) if the white area near one of the edges appears right next to a red (blue) region.

Taking into account equation \eqref{57-bis} we observe some remarkable behaviors:
\begin{figure}[htbp]
	\begin{center}
		\includegraphics[width=0.3\linewidth]{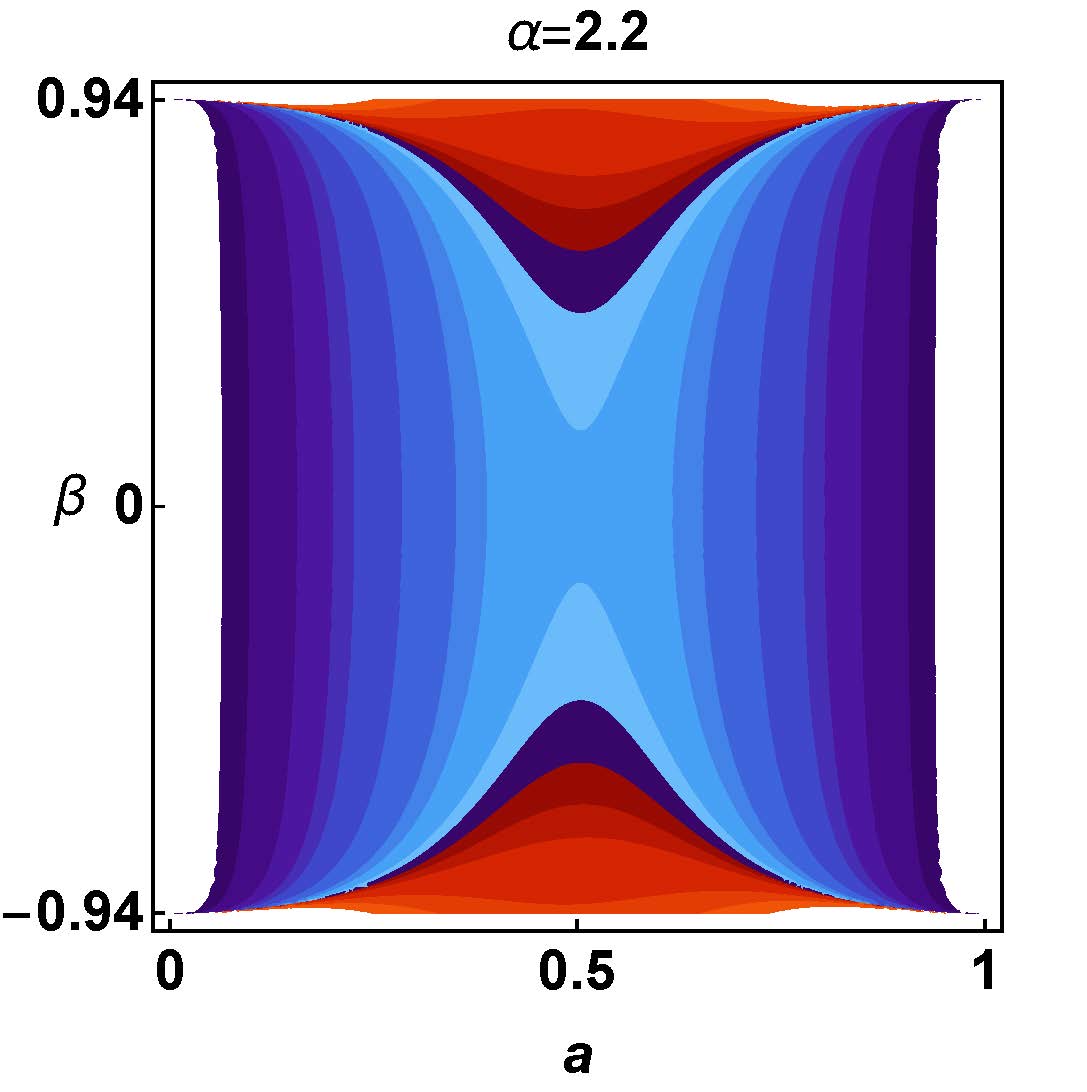}\quad \includegraphics[width=0.3\linewidth]{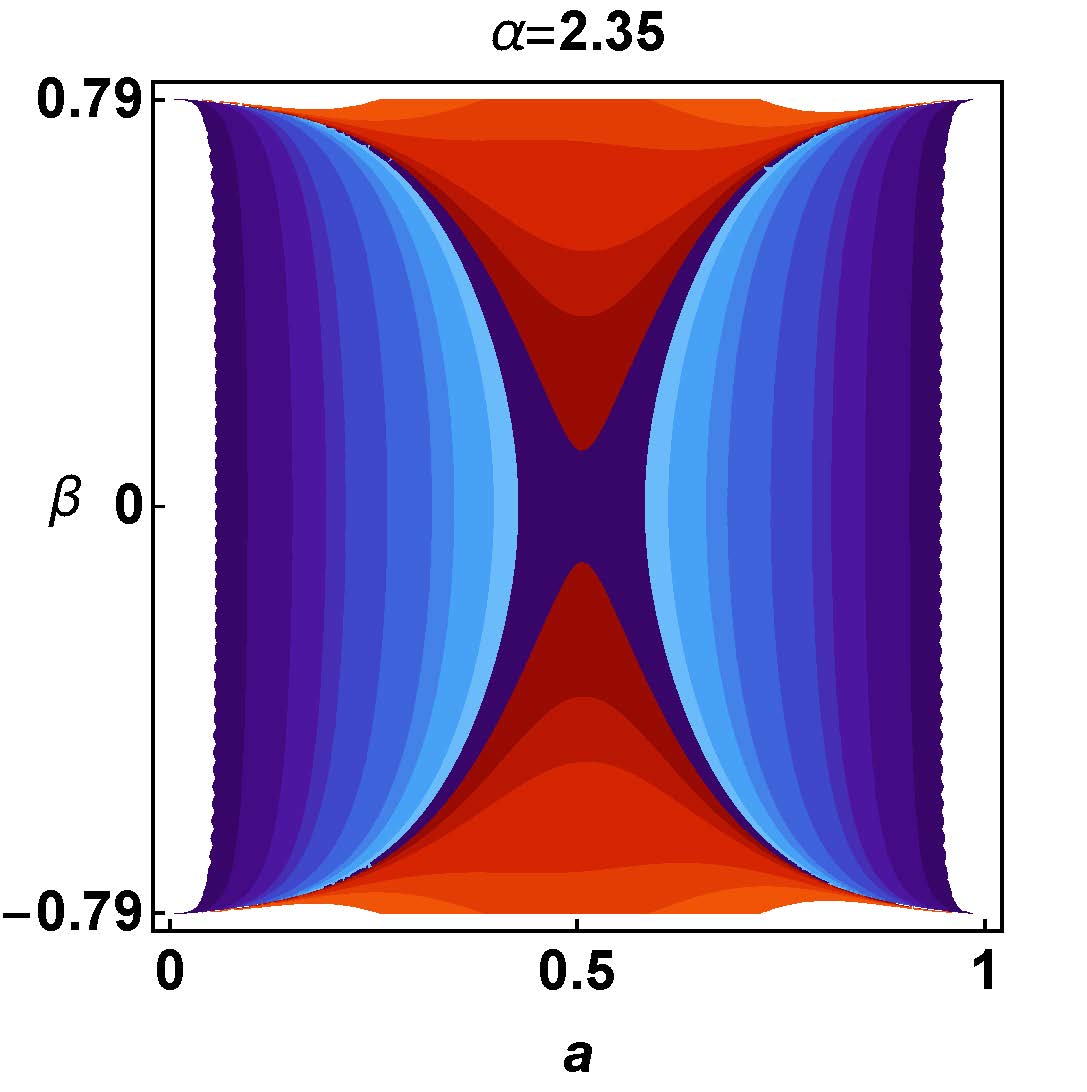} \quad \includegraphics[width=0.3\linewidth]{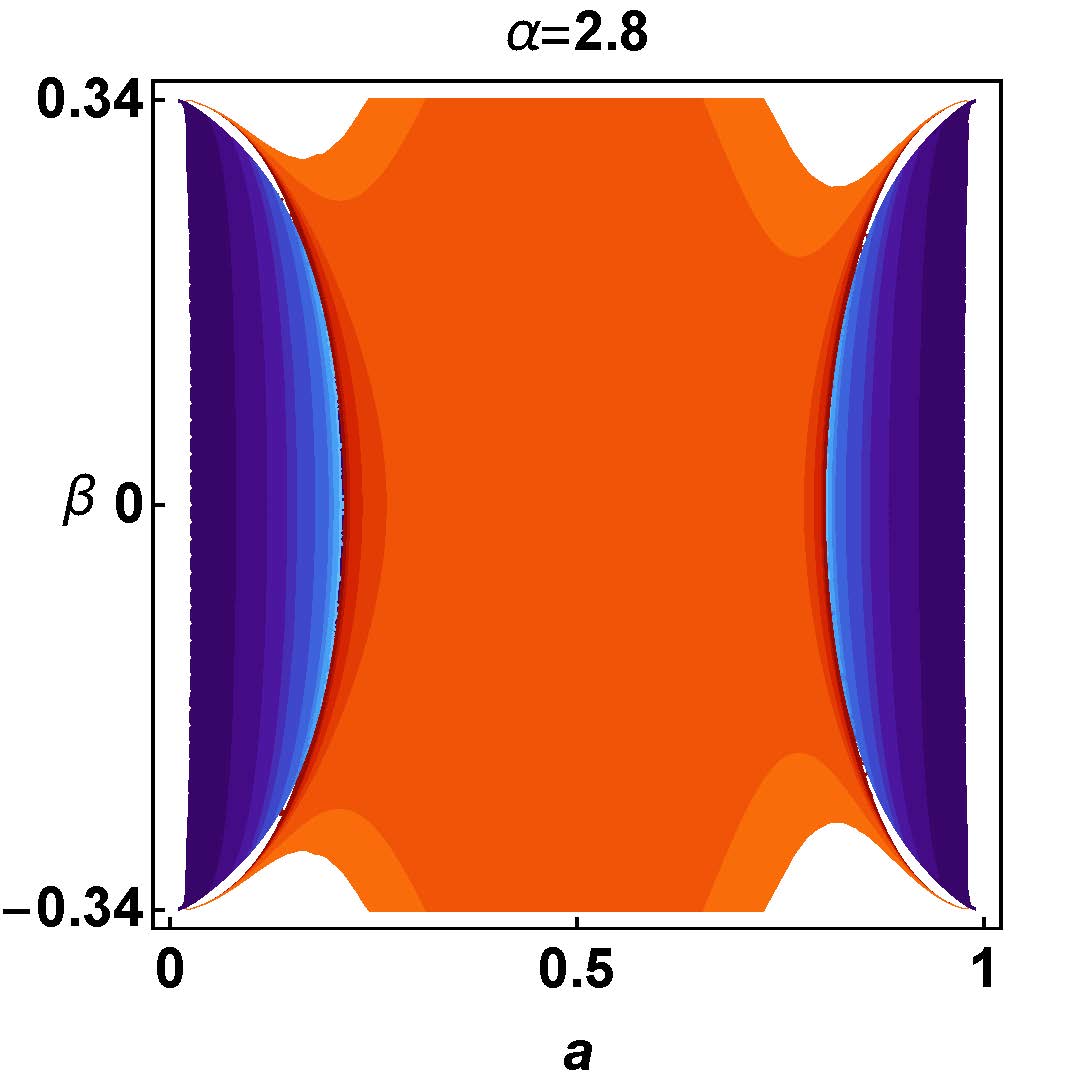}
		\caption{(color online) Behavior of the Casimir force \eqref{57-bis} as a function of the parameter $\beta$ of the piston characterised by ${\bf U}_1$ and the position $a$ of the piston, for different values of $\alpha$. The rest of the parameters are fixed to $L=1$, $\theta=\gamma=\pi/2$, $n_1=q_3=1$, and $\gamma=0$. The curves separating positive force (red color scale) and negative force (blue color scale) correspond to zero Casimir force situations.}
		\label{fig-3}
	\end{center}
\end{figure}
\begin{enumerate}
\item In all cases we are considering, except for the ones in Fig. \ref{fig-3}, we have that $n_{3}=q_{3}=0$.
It is not difficult to realize that for $n_{3}=q_{3}=0$, the function $h_{\lambda}(k,a;S,{\bf U}_1)$ is proportional to $\cos[k(2a-L)]$, and, hence, is an even function with respect to the midpoint $a=L/2$. Obviously the Casimir force, being the $k$-integral of the logarithmic derivative of $h_{\lambda}(k,a;S,{\bf U}_1)$, is an odd function with respect to the midpoint $a=L/2$. This implies, in particular, that in these cases the Casimir force is always zero at, at least, $a=L/2$. Let us point out that the force can vanish at other points of the interval, however these points of vanishing Casimir energy need to appear in pairs which are symmetric with respect to $a=L/2$. This behavior can be clearly observed from the graphs. For the cases in Fig. \ref{fig-3} we have, instead, $n_{3}=0$, $q_{3}=1$, and $\gamma=\theta=\pi/2$. In these cases the function $h_{\lambda}(k,a;S,{\bf U}_1)$ becomes proportional to $\sin[k(2a-L)]$. By following the argument outlined in the previous paragraph, the Casimir force is, then, an even function of $a$ with respect to $a=L/2$.
This means that the Casimir force at $a=L/2$ does not have to be necessarily zero.
\begin{figure}[htbp]
	\begin{center}
		\includegraphics[width=0.3\linewidth]{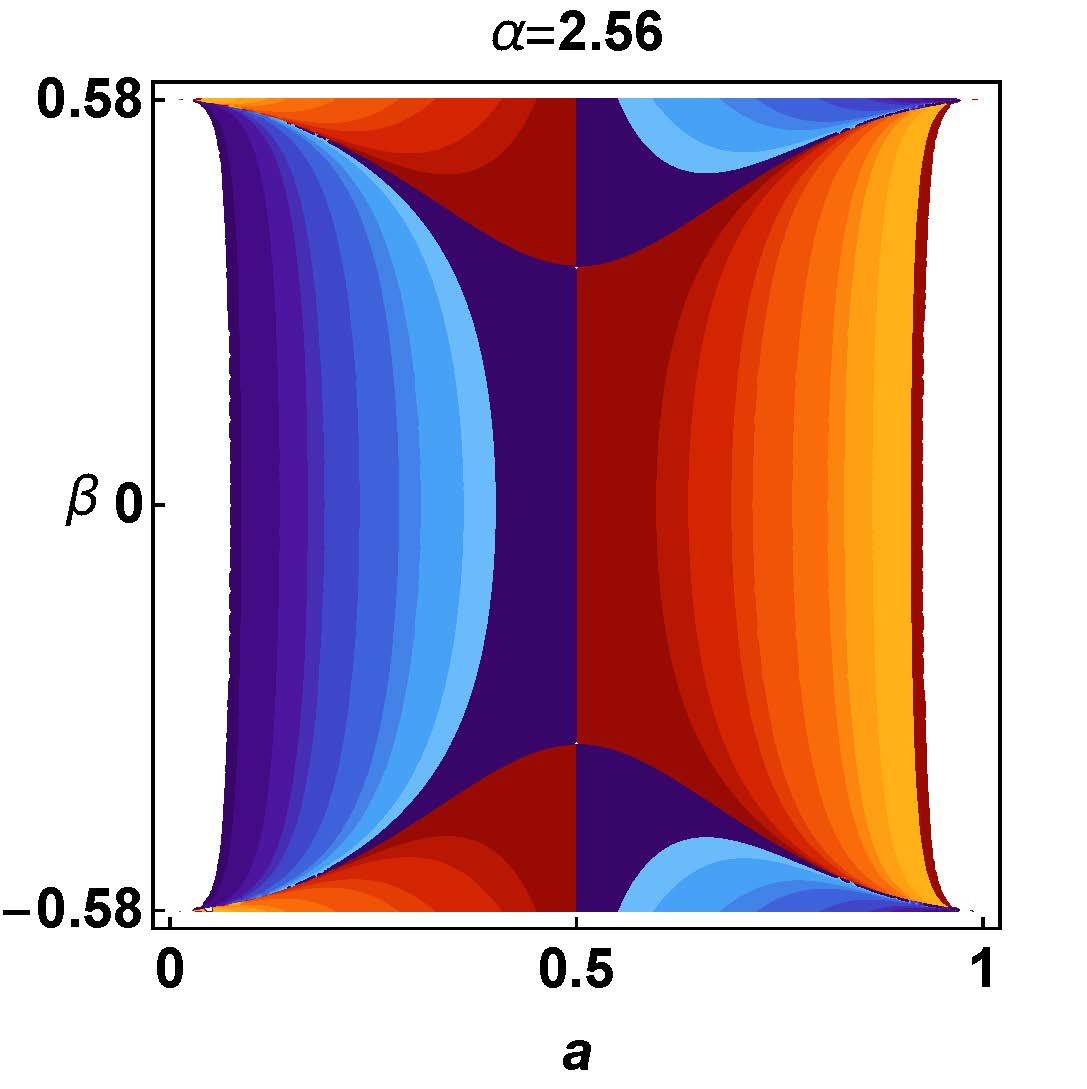}\quad \includegraphics[width=0.3\linewidth]{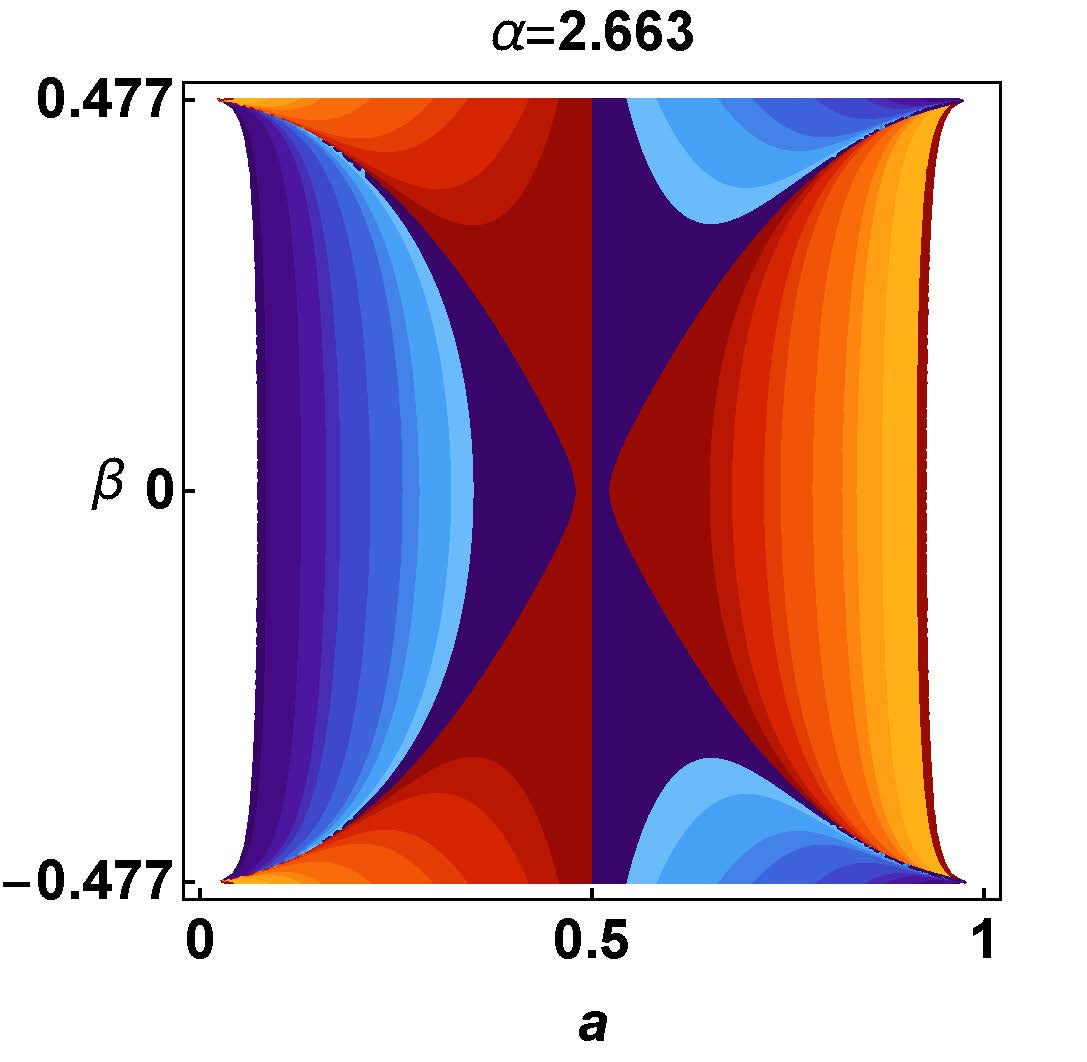} \quad \includegraphics[width=0.3\linewidth]{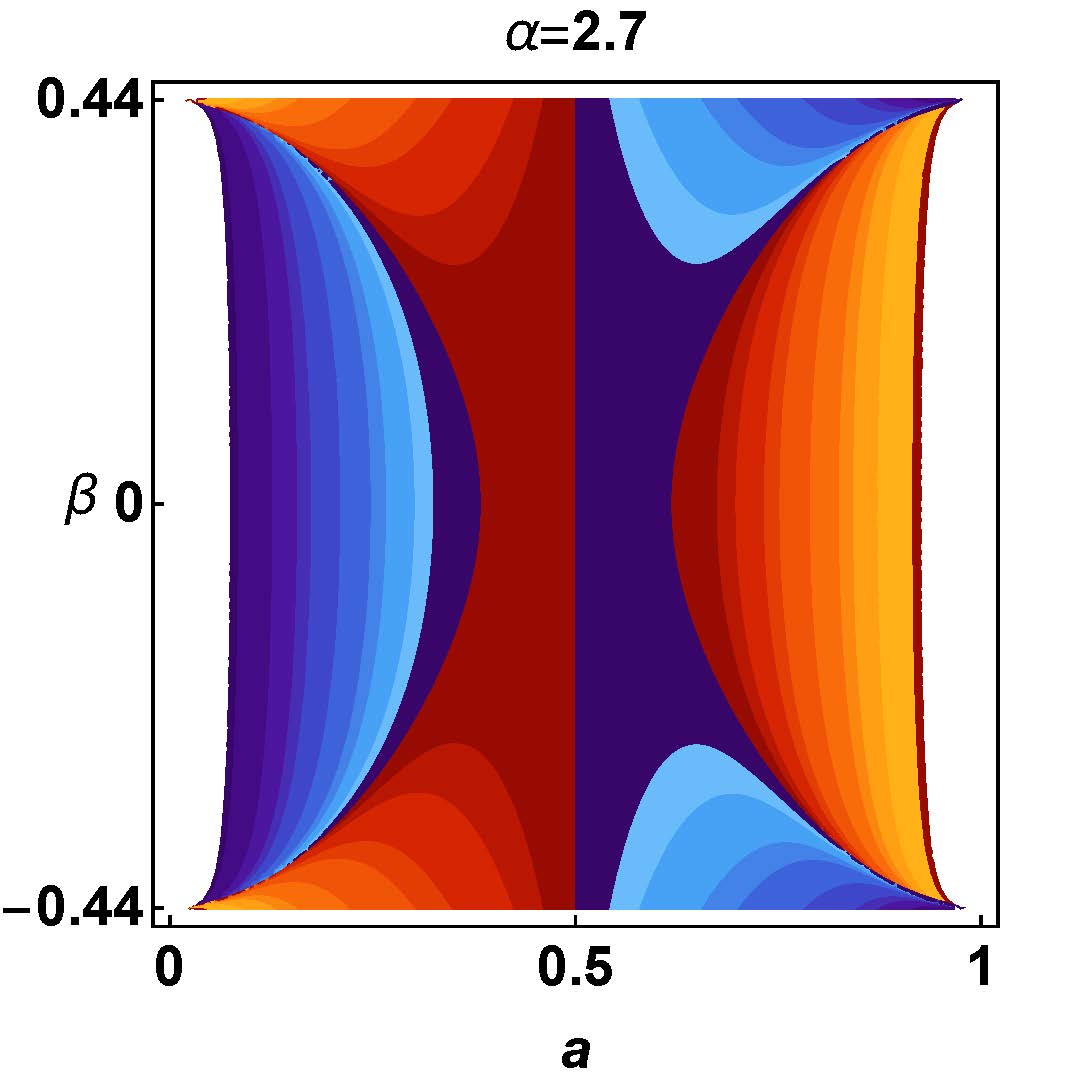}
		\caption{(color online) Behavior of the Casimir force \eqref{57-bis} as a function of the parameter $\beta$ of the piston characterised by ${\bf U}_1$ and the position $a$ of the piston, for different values of $\alpha$. The rest of the parameters are fixed to $L=1$, $\theta=1.5$, $\gamma=0$, $n_1=q_2=1$, and $\gamma=0$. The curves separating positive force (red color scale) and negative force (blue color scale) correspond to zero Casimir force situations.}
		\label{fig-4}
	\end{center}
\end{figure}
\item There exist regions in the space of free parameters for which the resulting Casimir force on the piston is either non-negative or non-positive for all values of the position $a$. In these situations the Casimir force will tend to move the piston to the right edge (if the force is non-negative) or to the left edge (if the force is non-positive).
An example of this behavior can be seen, for instance, in the first plot of Fig. \ref{fig-3}. For $\beta=0$ the force is always negative and, hence, the piston is moved towards the left edge.
In the situation we are considering, if any points of zero force are present, they would represent points of unstable equilibrium for the piston.

\item In Figs. \ref{fig-2}, \ref{fig-4} and \ref{fig-5} the Casimir force is, as explained earlier, an odd function of $a$ with respect to $a=L/2$. In these situations the points of vanishing force, which necessarily exist, can be points of either stable or unstable equilibrium. Let $\epsilon >0$. If $a_{0}$ is a point for which $F_{\textrm{Cas}}(a_{0})=0$,
then $a_{0}$ is a point of \emph{stable} equilibrium for the piston if $F_{\textrm{Cas}}(a_{0}-\epsilon)>0$ and $F_{\textrm{Cas}}(a_{0}+\epsilon)<0$. On the other hand, if
$F_{\textrm{Cas}}(a_{0}-\epsilon)<0$ and $F_{\textrm{Cas}}(a_{0}+\epsilon)>0$ then $a_{0}$ is a point of \emph{unstable} equilibrium for the piston.
Since the Casimir force in Figs. \ref{fig-2}, \ref{fig-4} and \ref{fig-5} is an odd function of $a$, then we must have an odd number of points where the force vanishes. These points of stable and unstable equilibrium must alternate as it can be clearly seen in the graphs.
\begin{figure}
\begin{center}
\includegraphics[width=0.335\linewidth]{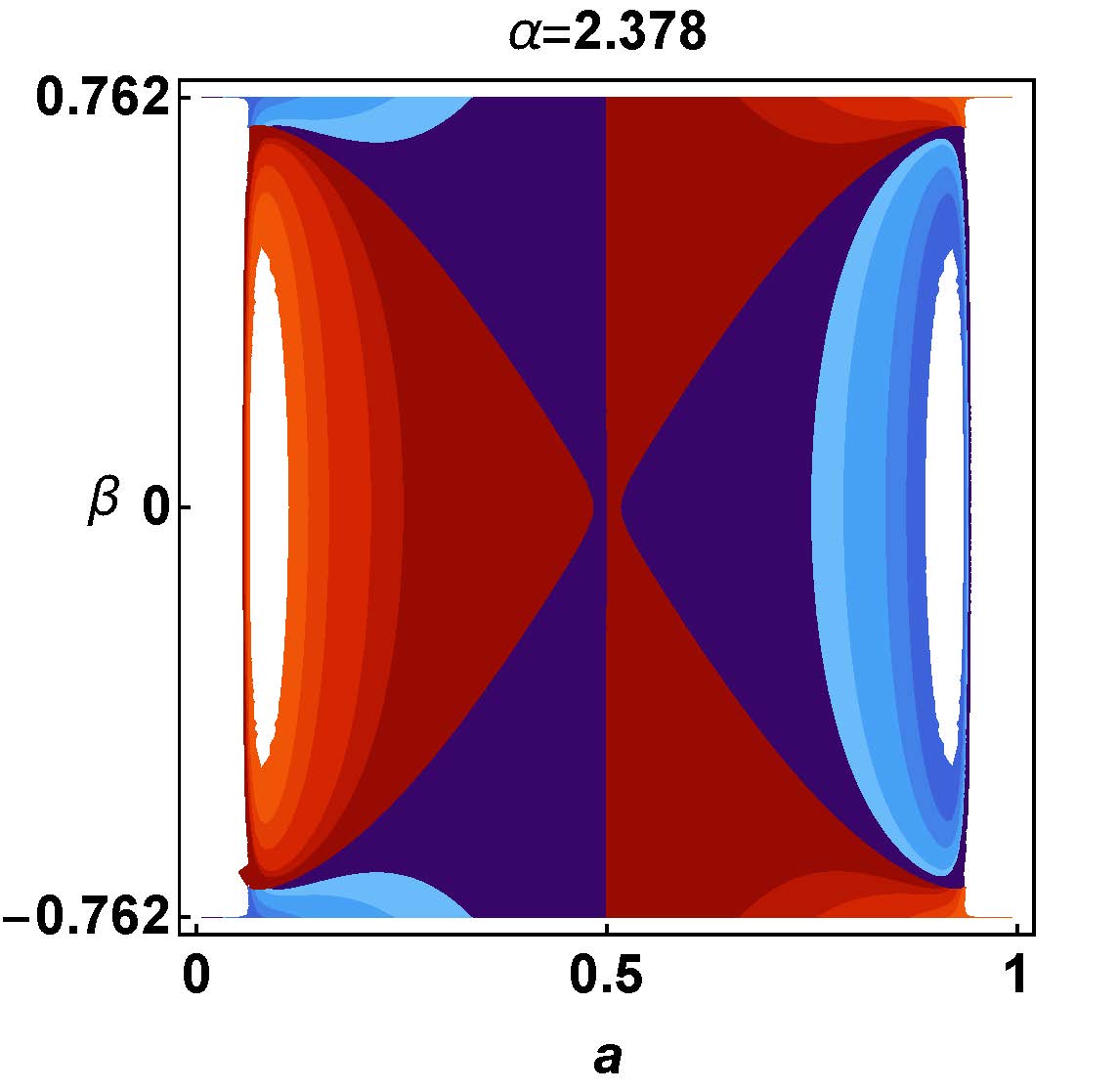}
\caption{(color online) Behavior of the Casimir force \eqref{57-bis} as a function of the parameter $\beta$ of the piston characterised by ${\bf U}_1$ and the position $a$ of the piston. The rest of the parameters are fixed to $L=1$, $\alpha=2.378$, $\theta=2$, $\gamma=1.14$, $n_1=q_2=1$, and $\gamma=0$. The curves separating positive force (red color scale) and negative force (blue color scale) correspond to zero Casimir force situations.}
\label{fig-5}
\end{center}
\end{figure}
\end{enumerate}

\subsection{Disk}

As a further example we consider the transverse manifold $N$ to be a disk of unit radius. The eigenfunctions of the Laplacian $\Delta_{N}$ can be found by using separation
of variables once $\Delta_{N}$ is written in polar coordinates $(r,\vartheta)$. By imposing periodicity of the solution with respect to the angular variable $\vartheta$
and Dirichlet boundary conditions at $r=1$, the eigenvalues can be easily found to be $\lambda_{kn}^{2}$ which can be determined as the zeroes of the Bessel function of the first kind
\begin{equation}\label{58}
J_{n}(\lambda_{kn})=0\;.
\end{equation}
One can show that the degeneracy of the eigenvalues satisfies the relations $d(\lambda_{k0})=1$ and $d(\lambda_{kn})=2$ for $n\geq 1$.
The zeroes of the Bessel function of the first kind with integer order are well known and can be found in tables or with the help of a computer program.
The figures for the Casimir force look qualitatively the same as for the sphere presented in the previous subsection and we therefore do not include any more details.

\section{Concluding remarks}

In this work we have studied the Casimir energy and force for a scalar field propagating in a piston configuration of the type $I\times N$.
The field is constrained by boundary conditions that lead to a selfa-djoint boundary value problem for the Laplacian on the piston.
We have focused, here, primarily on all non-negative self-adjoint extensions that can be described by matrices in the subgroup $U(2)\times U(2)$ of $U(4)$.
In particular we have studied the most general boundary conditions that relate the edges $x=0$ and $x=L$ and the two opposite edges of the piston itself.
By using scattering theory we were able to find an expression whose zeroes implicitly determined the eigenvalues of the Laplacian with the general boundary conditions
considered. This secular equation has been used as a starting point of an integral representation for the spectral zeta function which was subsequently analytically continued
to a larger region of the complex plane. Moreover, the use of non-relativistic scattering theory enables one to understand the physics behind the Casimir force in terms of the well known non-relativistic scattering theory in one-dimension.
The Casimir energy associated with the piston configuration and the corresponding force have been computed by exploiting the analytically continued expression
of the spectral zeta function. The formula that we found for the Casimir energy and force is written in terms of the spectral zeta function associated with
the Laplacian on the transverse manifold $N$ and is valid for any $d$-dimensional compact Riemannian manifold $N$ with or without boundary.
We have found the Casimir energy for a piston configuration is, in general, not a well-defined quantity with the ambiguity depending
on the geometry of the manifold $N$. This aspect of the Casimir energy on piston configuration has already been observed in the literature
(see e.g. \cite{bordag09,fucci-npb15}). While the energy might not be well-defined, the force on the piston is free of ambiguities.
The general expression we obtained for the force allowed us to derive the graphs presented in the previous section for specific manifolds $N$ and a number of particular
boundary conditions. It is important to point out that our formula (\ref{57}) can be used to perform a numerical
analysis of the Casimir force for any suitable transverse manifold $N$ and for any allowed values of the parameters in ${\bf U}_{1}$ and ${\bf U}_{2}$ that characterize the boundary conditions.

\paragraph{Further comments}Although the restriction to the membrane configuration taken in this paper enabled us to study the Casimir force for a piston with significant generality, there is still more that can be investigated for this system. For instance, it would be very interesting to analyze the case in which the matrix of the general boundary condition \eqref{3} is allowed to be any element of $U(4)$. In this situation it is not possible to use the advantages provided by the quantum mechanical scattering theory for 1D systems. In addition enabling the general boundary condition to be given by an arbitrary element of $U(4)$ will provide much richer physical phenomena, since more bound states might appear, and the four boundaries will be completely entangled.

The study carried out in this paper can as well be extended to Dirac fields. In this case the separability of the problem is not possible in general, so the parallel and orthogonal modes are not completely decoupled, as one can see easily from Ref. \cite{donaire19}.

\subsection*{Acknowledgments}
JMMC and KK are grateful to the Spanish Government-MINECO (MTM2014- 57129-C2-1-P) for the financial support received. JMMC is grateful and the Junta de Castilla y Le\'on (BU229P18, VA137G18 and VA057U16) for the financial support. JMMC would like to deeply acknowledge and honour all the support, the teaching, and the knowledge received from Professor Jose M. Mu$\tilde{\rm n}$oz-Porras during all his life: {\it it is the greatest honour to be your son.}

\end{document}